\documentclass[
  final,
  3p,
  times
]{elsarticle}
\biboptions{sort&compress}

\journal{Annals of Physics}

\usepackage{amsmath,amssymb,mathtools,bm}
\usepackage{graphicx}
\usepackage{xcolor}
\usepackage[mathlines]{lineno}
\usepackage{nccmath}
\usepackage{dcolumn}
\usepackage{dsfont}
\usepackage{physics}
\usepackage{adjustbox}
\usepackage{comment}
\usepackage{upgreek}
\usepackage{hyperref}

\hypersetup{
  colorlinks=true,
  breaklinks=true,
  urlcolor=[rgb]{0,0,0.36},
  linkcolor=[rgb]{0,0,0.36},
  citecolor=[rgb]{0,0,0.36}
}

\usepackage[dvipsnames]{xcolor}

\begin{document}

\begin{frontmatter}

% \title{Bridging Linear and Nonlinear Spectroscopy: Bubble–Square Relations and Rephasing Echoes}

\title{From pair bubbles to rephasing echoes: nonlinear spectroscopy of pair continua}

% ---------- Affiliations ----------

\affiliation[eth]{
  organization={Institute for Theoretical Physics, ETH Zurich},
  city={Zurich},
  postcode={8093},
  country={Switzerland}
}

\affiliation[Oxford]{
  organization={Rudolf Peierls Centre for Theoretical Physics, University of Oxford},
  city={Oxford},
  postcode={OX1 3PU},
  country={United Kingdom}
}

% ---------- Authors ----------
\author[eth,Oxford]{Alex Gomez-Salvador\corref{cor1}}
\cortext[cor1]{Correspondence: \href{mailto:alex.gomezsalvador@physics.ox.ac.uk}{alex.gomezsalvador@physics.ox.ac.uk}}

\author[eth]{Eugene Demler}

\begin{abstract}
% Fractionalized quasiparticles often manifest in spectroscopy not as sharp resonances, but as broad continua of pair excitations. 
When a probe creates quasiparticles in pairs, the spread of their energies gives rise to a spectral continuum even when the individual excitations are sharp. Because linear response intertwines the distribution of pair energies with quasiparticle decay, two-dimensional spectroscopy has been proposed to disentangle them through rephasing pathways. In this work, we establish when and how the third-order response of bosonic and fermionic pair continua can be computed from generalized two-point functions, without a separate calculation of four-point correlation functions. Beyond simplifying the calculation of nonlinear response, these generalized two-point functions reveal how nonlinear spectroscopy accesses momentum and symmetry projections of the pair continuum that are inaccessible through the physical linear susceptibility. We develop an explicit computational framework for noninteracting quasiparticles and incorporate interactions between them within a ladder approximation. We analyze the resulting two-dimensional spectra in the rephasing and nonrephasing sectors, revealing how the formation of a collective mode and the associated redistribution of spectral weight manifest in nonlinear spectroscopy.
\end{abstract}

\end{frontmatter}

\section{Introduction}

Broad spectral continua can arise when a probe creates excitations in pairs, such as electron–hole pairs in semiconductors, Bogoliubov quasiparticles in superconductors, and pairs of magnons or fractionalized spinons in quantum magnets.
Linear spectroscopy reveals the distribution of spectral weight, but it generally does not determine whether the observed broadening originates from a spread of pair energies, homogeneous dephasing, or a combination of both. Two-dimensional nonlinear spectroscopy offers a way to disentangle these effects by resolving the response along multiple time and frequency axes. Such measurements can separate excitation pathways and reveal signatures of coherence, relaxation, interactions, and higher-order correlations that are difficult to access in conventional linear response~\cite{Mukamel,hamm_zanni_2011,Liu2025Qmat,fiore_2026_2D_diagrams,huang2025review,Huang2026}.

One particularly useful nonlinear pathway is the rephasing, or echo, peak, whose refocusing mechanism is closely analogous to the spin echoes introduced by Hahn~\cite{Hahn_1950}. In this process, the phase evolution acquired during the first time interval is reversed during the second, allowing reversible inhomogeneous dephasing to be distinguished from irreversible homogeneous decay. The resulting echo therefore provides a more discriminating probe of the microscopic mechanisms controlling quasiparticle coherence than standard linear-response protocols~\cite{Cundiff2013,SmallwoodCundiff2018}.  

Echo phenomena have been studied extensively, both theoretically and experimentally, in disordered many-body systems, particularly in inhomogeneous superconductors~\cite{Liu_2023_echo,salvador2025Echo_theory,salvador2025gapinhomogeneitycupratesview}. At the same time, multidimensional spectroscopy has been proposed as a probe of systems with fractionalized excitations~\cite{Wan2019SpinonEcho,Choi_2020,Li_2021,Spinon_Hart_2023,Potts_2024}. Analytical studies have treated several solvable or controlled settings, including Ising spinons, gapped spin liquids, Luttinger liquids, and weakly interacting fractionalized quasiparticles~\cite{Wan2019SpinonEcho,Choi_2020,Li_2021,Spinon_Hart_2023}, while numerical approaches based on tensor networks and exact diagonalization have addressed interacting spinons and confinement~\cite{Gao_2023,Watanabe_2024}. Related calculations have also considered Majorana excitations in Kitaev spin liquids and spinon dynamics in quantum spin ices~\cite{Qiang_2024,Potts_2024}.

Nonlinear coherent spectroscopy has long been recognized as a powerful probe of microscopic information beyond linear response, with more recent developments highlighting its potential for investigating many-body phenomena. In random quantum magnets, for instance, it can resolve critical behavior and relaxation-time distributions~\cite{parameswaran2020asymptotically1}. In gapped one-dimensional systems, quasiparticle interactions generate long-time singularities in the nonlinear response~\cite{fava2023divergent}, with corresponding signatures in two-dimensional coherent and pump--probe spectroscopy~\cite{fava2024longtimedivergencesnonlinearresponse}. More recently, its sensitivity to critical superconducting fluctuations has further demonstrated its potential as a probe of collective phenomena~\cite{liu2026probingphasefluctuationscuprates}.
The microscopic origin of this additional information depends crucially on how the probe couples to the excitations. For collective bosonic fields driven and measured linearly, the third-order response is governed by a connected four-point function and vanishes in a harmonic theory. A probe coupled bilinearly to the constituent fields can instead produce a finite third-order response even for noninteracting quasiparticles. In this paper we demonstrate that the connected response of the pair operator need not access independent higher-order correlators of its constituents, but rather distinct two-point correlation functions from that governing the linear susceptibility.

In this work, we further establish both the conditions under which the third-order response of a pair continuum can be reduced to two-point correlation functions and the procedure for carrying out this reduction, first for noninteracting quasiparticles and then for interacting pairs within a ladder approximation.
In Sec.~\ref{sec:bubble_square_relation}, we use a common causal Keldysh representation for bosonic and fermionic fields to relate the four-propagator square entering the third-order susceptibility to generalized pair bubbles. This relation allows one to obtain an analytical two-dimensional spectrum in terms of these two-point correlation functions. While the intensity along the rephasing ridge follows the weighted distribution of pair energies, its transverse width measures homogeneous pair dephasing. In contrast, the nonrephasing signal probes the frequency dependence of the generalized retarded bubble. In particular, its diagonal cut is directly proportional to the frequency derivative of said bubble.

The reduction of the third-order response to two-point correlation functions does not generally make the nonlinear signal reconstructible from the physical linear susceptibility. For a probe form factor $f_k$, the linear bubble carries the weight $|f_k|^2$, whereas the echo involves $|f_k|^4$. Different pulse polarizations can in principle generate further combinations of vertices, providing access to momentum and symmetry components that are not separately resolved in the linear measurement. Consider, as an illustrative example, a single-band superconductor with a mixed gap $\Delta(\phi)=\Delta_s+\Delta_d\cos 2\phi$. Polarization-dependent pair-breaking probes weight different regions of its quasiparticle continuum. For example, a $B_{1g}$ Raman vertex proportional to $\cos 2\phi$ enters the linear response through $\cos^2 2\phi$, whereas the corresponding quasiparticle echo contains $\cos^4 2\phi$. The echo therefore places greater relative weight on regions where the probe vertex is largest, providing a distinct angular projection of the continuum. Comparing linear and echo spectra across polarizations can consequently supply additional constraints on the gap anisotropy and the $s$--$d$ admixture.

In Sec.~\ref{sec:particle_particle_interactions}, we extend this construction to particle--particle interactions within a ladder approximation. The ladder corrections dress both the pair susceptibility and the external vertices of the nonlinear square while preserving the reduction to generalized two-point functions. This allows us to follow the transfer of spectral weight from an extended pair continuum into an isolated bound state only in terms of two-point functions. In the strong-binding regime, the response near the isolated pole reduces to that of an effective interacting collective field with a local quartic interaction inherited from the constituent square. We derive the corresponding rephasing and nonrephasing spectra and compare this effective description with the microscopic ladder-dressed response. Finally we provide our conclusions and outlook.

\section{Pair bubble--square relation for noninteracting pair continua}
\label{sec:bubble_square_relation}

\begin{figure}
    \centering
    \includegraphics[width=0.95\linewidth]{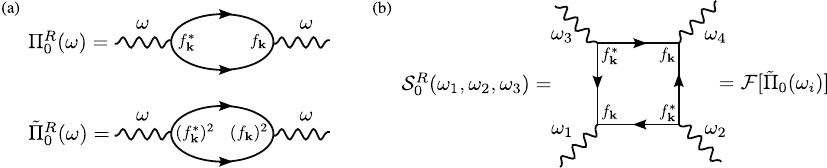}
    \caption{\textbf{Free diagrammatics for the pair-bubble and square.} (a) Pair-bubble diagrams corresponding to the linear susceptibility (top) and generalized pair bubble (bottom). (b) Pair-creation--annihilation square entering the third-order response. For noninteracting, well-defined quasiparticles within the on-shell approximation, the square diagram $\mathcal{S}_0^R(\omega_i)$ can be written solely in terms of the generalized pair bubble $\tilde{\Pi}_0(\omega_i)$.}
    \label{fig: pair_diagrams}
\end{figure}

In this section, we derive a relation between the pair bubble governing linear response, Fig.~\ref{fig: pair_diagrams}(a), and the four-propagator square entering the third-order response, Fig.~\ref{fig: pair_diagrams}(b), for noninteracting bosonic and fermionic quasiparticles. We focus on a single pair channel at vanishing external momentum, with momentum-dependent drive vertices $f_\mathbf{k}$ and an inversion-symmetric dispersion, $\xi_{-\mathbf{k}}=\xi_{\mathbf{k}}$. A uniform, frequency-independent damping rate $\gamma$ regularizes the quasiparticle propagators, and we require only that the equilibrium distribution function varies negligibly across each quasiparticle peak. 
% Using a common causal Keldysh representation, we first evaluate the pair bubble and then reduce the square diagram to a combination of form-factor-weighted two-point functions, obtaining the central generalized pair bubble–square relation in Eq.~\eqref{eqn:bubble_square_relation}. 
At the single-loop level, linear response is represented by two-vertex bubble diagrams, whereas the corresponding third-order nonlinear response involves four-vertex square diagrams \cite{Mahan2000,Mukamel1995}. As discussed below, this correspondence can extend beyond noninteracting theories, accommodating vertex corrections that preserve the relevant single-loop structure. Using a common causal Keldysh representation \cite{KamenevLevchenko2009, kamenev2023field}, we first evaluate the pair bubble and then reduce the square diagram to a combination of form-factor-weighted two-point functions, obtaining the central generalized pair bubble--square relation in Eq.~\eqref{eqn:bubble_square_relation}.
We subsequently apply this result to a two-pulse spectroscopy protocol and derive the analytical rephasing and nonrephasing spectra, Eqs.~\eqref{eqn:ABB_echo_map} and~\eqref{eqn:NR_ABB_map}. The former produces an echo ridge that resolves the pair spectral weight and homogeneous dephasing, while the latter provides a different probes of the retarded weighted pair bubble, reducing to its frequency derivative along the diagonal.

\subsection{A common causal representation for bosons and fermions}

We begin with a generic pair operator
\begin{equation}
P=\frac{1}{2}\psi^T\Lambda\psi,
\label{eqn: pair_def}
\end{equation}
where $\psi$ denotes the constituent quasiparticle field and $\Lambda$ acts on any internal spin, orbital, or flavor indices. Introducing
\begin{equation}
\eta=
\begin{cases}
+1, & \text{bosons},\\
-1, & \text{fermions},
\end{cases}
\end{equation}
the symmetry of the pair channel requires $\Lambda^T=\eta\Lambda$, and $\Lambda$ is symmetric for a bosonic pair and antisymmetric for a fermionic pair. To generate pair correlation functions, we couple $P$ to an external pair source $J$ on the two branches of the Keldysh contour, allowing the drive vertex to carry a momentum-dependent form factor $f_{\mathbf k}$,
\begin{equation}
S_J=-\sum_{s=\pm}s\int dt\sum_{\mathbf k}\left[J_s^\dagger f_{\mathbf k}P_{s,\mathbf k}+J_sf_{\mathbf k}^*P_{s,\mathbf k}^\dagger\right],\qquad P_{s,\mathbf k}=\frac12\psi_{s,\mathbf k}^T\Lambda\psi_{s,-\mathbf k}.
\end{equation}
Here $P_{s,\mathbf k}$ denotes the momentum-resolved pair bilinear at zero total momentum, while the form factor belongs only to the drive vertex. Since $P$ is a bosonic operator for both statistics, the source is rotated in the usual classical--quantum basis,
\begin{equation}
J_c=\frac{J_++J_-}{\sqrt2},
\qquad
J_q=\frac{J_+-J_-}{\sqrt2}.
\end{equation}
The constituent fields require slightly different rotations. For bosons,
\begin{equation}
\psi_c=\frac{\psi_++\psi_-}{\sqrt2},
\qquad
\psi_q=\frac{\psi_+-\psi_-}{\sqrt2},
\qquad
\bar\psi_c=\frac{\bar\psi_++\bar\psi_-}{\sqrt2},
\qquad
\bar\psi_q=\frac{\bar\psi_+-\bar\psi_-}{\sqrt2}.
\end{equation}
We write
\begin{equation}
\Psi_B=
\begin{pmatrix}
\psi_c\\
\psi_q
\end{pmatrix},
\qquad
\bar\Psi_B=
\begin{pmatrix}
\bar\psi_c&\bar\psi_q
\end{pmatrix}.
\end{equation}
For fermions we use the conventional Larkin--Ovchinnikov rotation,
\begin{equation}
\psi_1=\frac{\psi_++\psi_-}{\sqrt2},
\qquad
\psi_2=\frac{\psi_+-\psi_-}{\sqrt2},
\qquad
\bar\psi_1=\frac{\bar\psi_+-\bar\psi_-}{\sqrt2},
\qquad
\bar\psi_2=\frac{\bar\psi_++\bar\psi_-}{\sqrt2}.
\end{equation}
Note the different transformation for the barred Grassmann fields. We correspondingly define
\begin{equation}
\Psi_F=
\begin{pmatrix}
\psi_1\\
\psi_2
\end{pmatrix},
\qquad
\bar\Psi_F=
\begin{pmatrix}
\bar\psi_1&\bar\psi_2
\end{pmatrix}.
\end{equation}

Despite the different rotation of the barred fermionic fields, the pair vertices take the same form for bosons and fermions. Substituting the transformations above into the pair source and using $\Lambda^T=\eta\Lambda$, one obtains in both cases
\begin{equation}
S_J=-\frac{1}{2\sqrt2}\int dt\sum_{\mathbf k}\Big[
f_{\mathbf k}J_q^\dagger\Psi_{\mathbf k}^T(\sigma_0\otimes\Lambda)\Psi_{-\mathbf k}
+f_{\mathbf k}J_c^\dagger\Psi_{\mathbf k}^T(\sigma_x\otimes\Lambda)\Psi_{-\mathbf k}
+f_{\mathbf k}^*J_q\bar\Psi_{-\mathbf k}(\sigma_0\otimes\Lambda^\dagger)\bar\Psi_{\mathbf k}^T
+f_{\mathbf k}^*J_c\bar\Psi_{-\mathbf k}(\sigma_x\otimes\Lambda^\dagger)\bar\Psi_{\mathbf k}^T
\Big].
\label{eqn: S_drive}
\end{equation}
The Pauli matrices act in Keldysh space, while $\Lambda$ acts in the internal pair space. From this point onward, we restrict to a single fixed pair channel, such as identical bosons or fermionic quasiparticles with opposite spins and the same dispersion, and suppress its internal vertex structure.

The difference between bosonic and fermionic statistics remains in the conventional single-particle Green's-function matrices. We generally have
\begin{equation}
\check G_B(\omega)
=
\begin{pmatrix}
G^K(\omega)&G^R(\omega)\\
G^A(\omega)&0
\end{pmatrix},
\qquad 
\text{and}
\qquad
\check G_F(\omega)
=
\begin{pmatrix}
G^R(\omega)&G^K(\omega)\\
0&G^A(\omega)
\end{pmatrix},
\end{equation}
for bosons and fermions respectively. Here $G^{R/A/K}$ correspond to the retarded $(R)$, advanced $(A)$ and Keldysh $(K)$ components of the Green's function respectively. It is desirable to remove this representational difference between bosons and fermions by defining the causal matrix
\begin{equation}
\mathbb G(\omega)
\equiv
\begin{pmatrix}
G^R(\omega)&G^K(\omega)\\
0&G^A(\omega)
\end{pmatrix}=
\begin{cases}
\check G_B(\omega)\sigma_x, & \text{bosons},\\
\check G_F(\omega), & \text{fermions}.
\end{cases}
\label{eqn:causal_G}
\end{equation}
The second propagator in a pair loop has the opposite frequency orientation, as we will see. We therefore introduce analogously
\begin{equation}
\overline{\mathbb G}(\omega)
\equiv
\sigma_x\mathbb G^T(\omega)\sigma_x
=
\begin{pmatrix}
G^A(\omega)&G^K(\omega)\\
0&G^R(\omega)
\end{pmatrix}=
\begin{cases}
\check G_B^T(\omega)\sigma_x, & \text{bosons},\\
\sigma_x\check G_F^T(\omega)\sigma_x, & \text{fermions}.
\end{cases}
\label{eqn:causal_Gbar}
\end{equation}
The advantage of Eqs.~\eqref{eqn:causal_G} and \eqref{eqn:causal_Gbar} is that all subsequent expressions can now be written in exactly the same form for bosonic and fermionic quasiparticles.

We consider quasiparticles with a frequency-independent damping rate $\gamma$, governed by the retarded and advanced propagators
\begin{equation}
G_{\mathbf k}^{R/A}(\nu)
=
\frac{1}{\nu-\xi_{\mathbf k}\pm i\gamma}.
\label{eqn: free_propagator}
\end{equation}
In thermal equilibrium, the Keldysh component of the Green's function obeys
\begin{equation}
G_{\mathbf k}^K(\nu)
=
F_\eta(\nu)
\left[
G_{\mathbf k}^R(\nu)-G_{\mathbf k}^A(\nu)
\right],
\qquad \text{where} \qquad
F_\eta(\nu)
=
\begin{cases}
\coth\left(\dfrac{\beta\nu}{2}\right)=1+2n_B(\nu), & \eta=+1,\\
\tanh\left(\dfrac{\beta\nu}{2}\right)=1-2n_F(\nu), & \eta=-1.
\end{cases}
\label{eqn:statistics_F}
\end{equation}
We assume that the quasiparticles are sufficiently long lived such that their linewidth $\gamma$ is small compared with the frequency scale over which the equilibrium distribution function $F_\eta$ varies. The single-particle spectral function is then sharply concentrated near the quasiparticle pole $\nu=\xi_{\mathbf k}$, allowing $F_\eta$ to be evaluated on shell:
\begin{equation}
F_\eta(\nu)\simeq F_{\eta,\mathbf k}\equiv F_\eta(\xi_{\mathbf k}),
\qquad
G_{\mathbf k}^K(\nu)
\simeq
F_{\eta,\mathbf k}
\left[
G_{\mathbf k}^R(\nu)-G_{\mathbf k}^A(\nu)
\right],
\label{eqn: narrow_linewidth}
\end{equation}
and this ``on-shell'' approximation is therefore controlled by
\begin{equation}
\gamma
\left|
\partial_\omega F_\eta(\omega)
\right|_{\omega={\xi_{\mathbf k}}}
\ll
|F_{\eta,\mathbf k}|.
\end{equation}
At zero temperature, $F_\eta(\omega)$ becomes frequency independent on each fixed-sign frequency branch, so the approximation becomes exact away from $\omega=0$ in the zero linewidth limit. For a finite linewidth this remains accurate when the peak is sufficiently far from $
\omega=0$. Importantly, this is only a quasiparticle or narrow-linewidth approximation and no assumption is made about the ratio of $\gamma$ to the external frequency differences. Within this approximation, the causal matrix takes the form
\begin{equation}
\mathbb G_{\mathbf k}(\omega)
=
\begin{pmatrix}
G_{\mathbf k}^R(\omega)
&
F_{\eta,\mathbf k}
\left[
G_{\mathbf k}^R(\omega)-G_{\mathbf k}^A(\omega)
\right]
\\
0
&
G_{\mathbf k}^A(\omega)
\end{pmatrix},
\end{equation}
and it admits the frequency-independent similarity transformation
\begin{equation}
\mathbb G_{\mathbf k}(\omega)
=
U_{\eta,\mathbf k}
\begin{pmatrix}
G_{\mathbf k}^R(\omega)&0\\
0&G_{\mathbf k}^A(\omega)
\end{pmatrix}
U_{\eta,\mathbf k}^{-1},
\qquad
U_{\eta,\mathbf k}
=
\begin{pmatrix}
1&-F_{\eta,\mathbf k}\\
0&1
\end{pmatrix}.
\label{eqn:similarity_G}
\end{equation}
and similarly for $\overline{\mathbb G}$, but with the roles of $U_{\eta,\mathbf k}$ and $U_{\eta,\mathbf k}^{-1}$ interchanged:
\begin{equation}
\overline{\mathbb G}_{-\mathbf k}(\omega)
=
U_{\eta,\mathbf k}^{-1}
\begin{pmatrix}
G_{-\mathbf k}^A(\omega)&0\\
0&G_{-\mathbf k}^R(\omega)
\end{pmatrix}
U_{\eta,\mathbf k}.
\label{eqn:similarity_Gbar}
\end{equation}
Since $U_{\eta,\mathbf k}$ is, within this ``on-shell'' approximation, frequency independent, the matrices $\mathbb G(\omega)$ at different frequencies commute with one another, and the same holds for $\overline{\mathbb G}(\omega)$:
\begin{equation}
[\mathbb G_{\mathbf k}(\omega),\mathbb G_{\mathbf k}(\omega')]=0,
\qquad
[\overline{\mathbb G}_{\mathbf k}(\omega),\overline{\mathbb G}_{\mathbf k}(\omega')]=0.
\label{eqn:same_orientation_commutators}
\end{equation}
Note that for systems where $\xi_{\mathbf k}=\xi_{-\mathbf k}$, like the ones under study here, we also have
\begin{equation}
[\mathbb G_{\mathbf k}(\omega),\mathbb G_{-\mathbf k}(\omega')]=0,
\qquad
[\overline{\mathbb G}_{\mathbf k}(\omega),\overline{\mathbb G}_{-\mathbf k}(\omega')]=0.
\label{eqn:same_orientation_commutators_k}
\end{equation}
This commutative structure is what ultimately allows the square diagram to be reduced to pair bubbles.

\subsection{The pair susceptibility }

We first define the reduced retarded pair susceptibility, see Fig.~\ref{fig: pair_diagrams}(a). With one quantum and one classical pair vertex, Eq.~\eqref{eqn: S_drive} gives, in either the bosonic or fermionic representation,
\begin{equation}
\Pi_0^R(Q)
=
i\int_P|f_{\mathbf p-\mathbf q/2}|^2
\operatorname{Tr}_{\mathrm K}
\left[
\check G(P)\sigma_x\check G^T(Q-P)
\right],
\end{equation}
where $P=(\nu,\mathbf p)$, and $Q=(\omega,\mathbf q)$. We use $\Pi_0^{R/A}$ to denote reduced loop functions in which overall numerical prefactors arising from combinatorics, statistics, and vertex conventions are suppressed, while retaining the standard conjugation relation $\Pi_0^A(\omega) = [\Pi_0^R(\omega)]^*$. In terms of the common causal matrices, see Eqs.~\eqref{eqn:causal_G} and~\eqref{eqn:causal_Gbar}, the pair loop can then be written in the same form for both statistics as
\begin{equation}
\Pi_0^R(Q)
=
i\int_P|f_{\mathbf p-\mathbf q/2}|^2
\operatorname{Tr}_{\mathrm K}
\left[
\sigma_x
\mathbb G(P)
\overline{\mathbb G}(Q-P)
\right],
\label{eqn:universal_pair_bubble_matrix}
\end{equation}
At zero external momentum we have, after computing the trace over the Keldysh space,
\begin{equation}
\Pi_0^R(\omega)
= i
\int_{\mathbf k}|f_{\mathbf k}|^2
\int\frac{d\nu}{2\pi}
\left[
G_{\mathbf k}^K(\nu)
G_{-\mathbf k}^R(\omega-\nu)
+
G_{\mathbf k}^R(\nu)
G_{-\mathbf k}^K(\omega-\nu)
\right].
\end{equation}
Using the fact that the mixed $RA$ and $AR$ integrals vanish, while $\int d\nu\,
G_{\mathbf k}^R(\nu)
G_{-\mathbf k}^R(\omega-\nu)/2\pi
=
-i/
(\omega-2\xi_{\mathbf k}+2i\gamma$),
we recover the standard non-interacting pair-bubble
\begin{equation}
\Pi_0^{R/A}(\omega)
=
2
\int_{\mathbf k}
\frac{|f_{\mathbf k}|^2F_{\eta,\mathbf k}}
{\omega-2\xi_{\mathbf k}\pm2i\gamma}.
\label{eqn: bare_bubble}
\end{equation}
The two-particle damping rate is therefore $2\gamma$, corresponding to the sum of the damping rates of the two constituent quasiparticles. For future reference,
\begin{equation}
\Pi_0^R(\omega)-\Pi_0^A(\omega)
=
-4\pi i
\int_{\mathbf k}|f_{\mathbf k}|^2
F_{\eta,\mathbf k}
\delta_{2\gamma}(\omega-2\xi_{\mathbf k}),
\label{eqn:finite_gamma_spectral_jump}
\qquad
\text{where}
\qquad
\delta_{2\gamma}(s)
=
\frac{1}{\pi}
\frac{2\gamma}{s^2+4\gamma^2}.
\end{equation}
%
% By contrast, as we will see next, the square diagram contains four external drive vertices and therefore carries the momentum weight $|f_{\mathbf k}|^4$. It is useful to define the generalized bubble
% \begin{equation}
% \tilde{\Pi}_0^{R/A}(\omega)=2\int_{\mathbf k}|f_{\mathbf k}|^4\frac{F_{\eta,\mathbf k}}{\omega-2\xi_{\mathbf k}\pm2i\gamma}.
% \label{eqn:weighted_bubble}
% \end{equation}
% At vanishing external momentum, the internal momentum $\mathbf k$ is conserved around the loop. Consequently, the momentum-dependent form factors do not interfere with the frequency manipulations underlying the bubble--square reduction. For $f_{\mathbf k}=1$, $\tilde\Pi_0=\Pi_0$.

\subsection{The square diagram and its reduction}

We next turn to the four-propagator loop that generates the corresponding contribution to the third-order response, see Fig.~\ref{fig: pair_diagrams}(b). In contrast to the pair bubble, the square contains several external frequency insertions, and its reduction depends sensitively on how these frequencies are routed through the internal propagators. It is therefore useful to fix the frequency convention explicitly before evaluating the Keldysh structure. We take four external frequencies satisfying
\begin{equation}
\omega_1+\omega_4=\omega_2+\omega_3,
\label{eqn: frequency_conservation}
\end{equation}
with legs $1$ and $4$ outgoing and legs $2$ and $3$ incoming, see Fig.~\ref{fig: pair_diagrams}(b). Starting with loop frequency $\nu$ on the line between legs $1$ and $2$, the four propagators carry frequencies
\begin{equation}
y=\nu,
\qquad
v=\omega_2-\nu,
\qquad
x=\nu+\omega_3-\omega_1,
\qquad
u=\omega_1-\nu.
\label{eqn:frequency_routing}
\end{equation}
The corresponding reduced square can be written in the conventional bosonic or fermionic basis as
\begin{equation}
\mathcal S_0^R
= i\int_{\mathbf{k},\nu}|f_{\mathbf k}|^4
\operatorname{Tr}_{\mathrm K}
\left[
\check G_{\mathbf k}(x)\sigma_x
\check G_{-\mathbf k}^T(u)\sigma_x
\check G_{\mathbf k}(y)\sigma_x
\check G_{-\mathbf k}^T(v)
\right],
\end{equation}
where we have introduced the notation $    \int_{\mathbf{k},\nu} = \int_{\mathbf k}\int\frac{d\nu}{2\pi}$. Using Eqs.~\eqref{eqn:causal_G} and \eqref{eqn:causal_Gbar}, both statistics reduce to the same expression
\begin{equation}
\mathcal S_0^R=i\int_{\mathbf k,\nu}|f_{\mathbf k}|^4\operatorname{Tr}_{\mathrm K}\left[\sigma_x\mathbb G_{\mathbf k}(x)\overline{\mathbb G}_{-\mathbf k}(u)\mathbb G_{\mathbf k}(y)\overline{\mathbb G}_{-\mathbf k}(v)\right]
=i\int_{\mathbf k,\nu}|f_{\mathbf k}|^4\operatorname{Tr}_{\mathrm K}\left[\sigma_x\mathbb G_x\overline{\mathbb G}_u\mathbb G_y\overline{\mathbb G}_v\right].
\label{eqn:universal_square}
\end{equation}
where in the second equality we have introduced the shorthand notation
\begin{equation}
    \mathbb G_x \equiv \mathbb G_\mathbf{k}(x), \qquad 
    \mathbb G_y \equiv \mathbb G_\mathbf{k}(y), \qquad
    \overline{\mathbb G}_u \equiv \overline{\mathbb G}_{-\mathbf{k}}(u), \qquad
    \overline{\mathbb G}_v \equiv \overline{\mathbb G}_{-\mathbf{k}}(v).
\end{equation}
Equations~\eqref{eqn:universal_pair_bubble_matrix} and \eqref{eqn:universal_square} are the starting point of the rest of this section. Our goal is to establish a relation between the linear and third-order nonlinear responses. The pair bubble contains a single product $\mathbb{G}\overline{\mathbb{G}}$, whereas the square diagram contains two such pairs with crossed frequency orderings, making a direct reduction appear nontrivial. We will show in the following, however, that under the stated ``on-shell'' approximation, the four-propagator structure can be decomposed into products of two-propagator contributions, and therefore into the computation of two-point functions.

The structure we seek to obtain from Eq.~\eqref{eqn:universal_square} is $\mathbb G_x\mathbb G_y\overline{\mathbb G}_u\overline{\mathbb G}_v$, so that propagators with the same frequency orientation appear adjacently. Their products can then be reduced to differences of single propagators using the identities derived below.  The only obstruction is commuting $\overline{\mathbb G}_u$ past $\mathbb G_y$. Direct multiplication gives
\begin{equation}
\left[
\overline{\mathbb G}_{-\mathbf k}(u),
\mathbb G_{\mathbf k}(y)
\right]
=
-2F_{\eta,\mathbf k}
\left[
G_{-\mathbf k}^R(u)-G_{-\mathbf k}^A(u)
\right]
\left[
G_{\mathbf k}^R(y)-G_{\mathbf k}^A(y)
\right]\begin{pmatrix}
0&1\\
0&0
\end{pmatrix}.
\label{eqn: cross_commutator}
\end{equation}
The commutator is therefore purely spectral, i.e., away from the spectral support given by $G^R-G^A\propto {\rm{Im}}\, G^R$ the matrices $\mathbb{G}$ and $\overline{\mathbb{G}}$ commute. We can therefore decompose $\mathcal S_0^R=\mathcal S_{0,\mathrm{reg}}^R+\mathcal S_{0,\mathrm{sp}}^R$, with
\begin{equation}
\mathcal S_{0,\mathrm{reg}}^R
= i
\int_{\mathbf k,\nu}|f_{\mathbf k}|^4
\operatorname{Tr}_{\mathrm K}
\left[
\sigma_x
\mathbb G_x
\mathbb G_y
\overline{\mathbb G}_u
\overline{\mathbb G}_v
\right],
\qquad
\text{and}
\qquad
\mathcal S_{0,\mathrm{sp}}^R
= i
\int_{\mathbf k,\nu}|f_{\mathbf k}|^4
\operatorname{Tr}_{\mathrm K}
\left[
\sigma_x
\mathbb G_x
\left[
\overline{\mathbb G}_u,\mathbb G_y
\right]
\overline{\mathbb G}_v
\right],
\end{equation}
where ``reg'' denotes the regular, all retarded, contribution and ``sp'' the spectral one containing the commutator. As we will see below, it is precisely the spectral term that gives rise to the echo peak and its associated rephasing properties.

\subsubsection{Reduction of the regular contribution}

We first consider the regular contribution and show that it can be reduced directly to a combination of retarded pair bubbles. To this end, we define the shorthand notation
\begin{equation}
\Delta_{31}\equiv\omega_3-\omega_1=x-y,
\qquad
\Delta_{21}\equiv\omega_2-\omega_1=v-u.
\end{equation}
Since the self-energy is frequency independent, propagators with the same analytic structure obey
\begin{equation}
G^{R/A}(z_1)G^{R/A}(z_2)
=
\frac{
G^{R/A}(z_2)-G^{R/A}(z_1)
}
{z_1-z_2}.
\end{equation}
Together with the similarity transformations in Eqs.~\eqref{eqn:similarity_G} and \eqref{eqn:similarity_Gbar}, this gives the matrix identities
\begin{equation}
\mathbb G_x\mathbb G_y
=
\frac{\mathbb G_y-\mathbb G_x}{\Delta_{31}},
\qquad
\overline{\mathbb G}_u\overline{\mathbb G}_v
=
\frac{\overline{\mathbb G}_u-\overline{\mathbb G}_v}{\Delta_{21}}.
\label{eqn:matrix_resolvents}
\end{equation}
Therefore
\begin{multline}
\mathcal S_{0,\mathrm{reg}}^R
=
\frac{i}{\Delta_{31}\Delta_{21}}
\int_{\mathbf k,\nu}|f_{\mathbf k}|^4
\operatorname{Tr}_{\mathrm K}
\left[
\sigma_x
(\mathbb G_y-\mathbb G_x)
(\overline{\mathbb G}_u-\overline{\mathbb G}_v)
\right]\\
=
\frac{i}{\Delta_{31}\Delta_{21}}
\int_{\mathbf k,\nu}|f_{\mathbf k}|^4
\operatorname{Tr}_{\mathrm K}
\left[
\sigma_x
\left(
\mathbb G_y\overline{\mathbb G}_u
-
\mathbb G_y\overline{\mathbb G}_v
-
\mathbb G_x\overline{\mathbb G}_u
+
\mathbb G_x\overline{\mathbb G}_v
\right)
\right].
\end{multline}
Each term is now precisely a weighted pair bubble of the form in Eq.~\eqref{eqn:universal_pair_bubble_matrix}, with $|f_{\mathbf k}|^2$ replaced by $|f_{\mathbf k}|^4$. Using the frequency convention defined in Eq.~\eqref{eqn:frequency_routing}, we obtain
\begin{equation}
\mathcal S_{0,\mathrm{reg}}^R
=
\frac{
\tilde{\Pi}_0^R(\omega_1)
-
\tilde{\Pi}_0^R(\omega_2)
-
\tilde{\Pi}_0^R(\omega_3)
+
\tilde{\Pi}_0^R(\omega_4)
}
{\Delta_{31}\Delta_{21}},
\label{eqn:regular_reduction}
\end{equation}
where we have defined the generalized pair bubble 
\begin{equation}
\tilde{\Pi}_0^{R/A}(\omega)=2\int_{\mathbf k}|f_{\mathbf k}|^4\frac{F_{\eta,\mathbf k}}{\omega-2\xi_{\mathbf k}\pm2i\gamma}.
\label{eqn:weighted_bubble}
\end{equation}
Note that for $f_{\mathbf k}=1$, $\tilde\Pi_0=\Pi_0$, recovering the linear pair susceptibility.

\subsubsection{Reduction of the spectral contribution}

We now turn to the spectral contribution. Unlike the regular term, its reduction involves products of retarded and advanced propagators, so the finite quasiparticle linewidth enters explicitly. Using Eq.~\eqref{eqn: cross_commutator}, the spectral contribution becomes
\begin{equation}
\begin{aligned}
\mathcal S_{0,\mathrm{sp}}^R
=
-2i
\int_{\mathbf k,\nu}|f_{\mathbf k}|^4
F_{\eta,\mathbf k}\,
G_{\mathbf k}^R(x)
\left[
G_{\mathbf k}^R(y)-G_{\mathbf k}^A(y)
\right]
\left[
G_{-\mathbf k}^R(u)-G_{-\mathbf k}^A(u)
\right]
G_{-\mathbf k}^R(v).
\end{aligned}
\label{eqn:spectral_start}
\end{equation}
At finite $\gamma$, the two spectral functions have a finite width and the remaining propagators cannot be evaluated at a single value of the internal frequency. However, the frequency-independent self-energy implies the mixed resolvent identities
\begin{equation}
G_{\mathbf k}^R(x)G_{\mathbf k}^A(y)
=
\frac{
G_{\mathbf k}^A(y)-G_{\mathbf k}^R(x)
}
{\Delta_{31}+2i\gamma},
\qquad
\text{and}
\qquad
G_{-\mathbf k}^A(u)G_{-\mathbf k}^R(v)
=
\frac{
G_{-\mathbf k}^A(u)-G_{-\mathbf k}^R(v)
}
{\Delta_{21}+2i\gamma},
\end{equation}
together with
\begin{equation}
G_{\mathbf k}^R(x)G_{\mathbf k}^R(y)
=
\frac{
G_{\mathbf k}^R(y)-G_{\mathbf k}^R(x)
}
{\Delta_{31}},
\qquad
\text{and}
\qquad
G_{-\mathbf k}^R(u)G_{-\mathbf k}^R(v)
=
\frac{
G_{-\mathbf k}^R(u)-G_{-\mathbf k}^R(v)
}
{\Delta_{21}},
\end{equation}
we obtain for the spectral part of the square diagram
\begin{equation}
\begin{aligned}
\mathcal S_{0,\mathrm{sp}}^R
=
-2 i
\int_{\mathbf k,\nu}|f_{\mathbf k}|^4
F_{\eta,\mathbf k}
\Bigg[
&
\frac{
\left[
G_{\mathbf k}^R(y)-G_{\mathbf k}^R(x)
\right]
\left[
G_{-\mathbf k}^R(u)-G_{-\mathbf k}^R(v)
\right]
}
{\Delta_{31}\Delta_{21}}
-
\frac{
\left[
G_{\mathbf k}^R(y)-G_{\mathbf k}^R(x)
\right]
\left[
G_{-\mathbf k}^A(u)-G_{-\mathbf k}^R(v)
\right]
}
{\Delta_{31}(\Delta_{21}+2i\gamma)}
\\
-&
\frac{
\left[
G_{\mathbf k}^A(y)-G_{\mathbf k}^R(x)
\right]
\left[
G_{-\mathbf k}^R(u)-G_{-\mathbf k}^R(v)
\right]
}
{(\Delta_{31}+2i\gamma)\Delta_{21}}
+
\frac{
\left[
G_{\mathbf k}^A(y)-G_{\mathbf k}^R(x)
\right]
\left[
G_{-\mathbf k}^A(u)-G_{-\mathbf k}^R(v)
\right]
}
{(\Delta_{31}+2i\gamma)(\Delta_{21}+2i\gamma)}
\Bigg].
\end{aligned}
\label{eqn:spectral_expanded}
\end{equation}
Each term can again be reduced to a pair bubble and we obtain
\begin{equation}
\begin{aligned}
\mathcal S_{0,\mathrm{sp}}^R
={}&
-
\frac{
\tilde{\Pi}_0^R(\omega_1)
-
\tilde{\Pi}_0^R(\omega_2)
-
\tilde{\Pi}_0^R(\omega_3)
+
\tilde{\Pi}_0^R(\omega_4)
}
{\Delta_{31}\Delta_{21}}
\\
&+
\frac{
\tilde{\Pi}_0^R(\omega_4)-\tilde{\Pi}_0^R(\omega_2)
}
{\Delta_{31}(\Delta_{21}+2i\gamma)}
+
\frac{
\tilde{\Pi}_0^R(\omega_4)-\tilde{\Pi}_0^R(\omega_3)
}
{(\Delta_{31}+2i\gamma)\Delta_{21}}
+
\frac{
\tilde{\Pi}_0^A(\omega_1)-\tilde{\Pi}_0^R(\omega_4)
}
{(\Delta_{31}+2i\gamma)(\Delta_{21}+2i\gamma)}.
\end{aligned}
\label{eqn:spectral_reduction}
\end{equation}
The first term cancels exactly against the regular contribution in Eq.~\eqref{eqn:regular_reduction}. The complete ordered square considering both terms can therefore be written entirely in terms of the weighted pair bubble in Eq.~\eqref{eqn:weighted_bubble},
\begin{equation}
\mathcal S_0^R(\omega_1,\omega_2,\omega_3)
=\frac{
\tilde{\Pi}_0^R(\omega_4)-\tilde{\Pi}_0^R(\omega_2)
}
{(\omega_3-\omega_1)(\omega_2-\omega_1+2i\gamma)}
+
\frac{
\tilde{\Pi}_0^R(\omega_4)-\tilde{\Pi}_0^R(\omega_3)
}
{(\omega_3-\omega_1+2i\gamma)(\omega_2-\omega_1)}
+
\frac{
\tilde{\Pi}_0^A(\omega_1)-\tilde{\Pi}_0^R(\omega_4)
}
{(\omega_3-\omega_1+2i\gamma)(\omega_2-\omega_1+2i\gamma)}.
\label{eqn:bubble_square_relation}
\end{equation}
Equation~\eqref{eqn:bubble_square_relation} is the main result of this section. It shows that the four-propagator loop entering $\chi^{(3)}$ can be rewritten entirely in terms of two-point functions. Once the generalized pair bubbles $\tilde{\Pi}_0$ are known, the square follows directly. Note that the apparent singularities at $\omega_3\to\omega_1$ or $\omega_2\to\omega_1$ are removable and that in the coincident-frequency limit, they reduce to frequency derivatives of $\tilde{\Pi}_0^{R/A}$. Within the on-shell approximation of Eq.~\eqref{eqn: narrow_linewidth} and for the frequency-independent self-energy considered here, Eq.~\eqref{eqn:bubble_square_relation} remains exact at finite $\gamma$. 
% In particular, no expansion in $\gamma/|\omega_i-\omega_j|$ has been made, so the relation remains valid when the external frequency differences are comparable to, or smaller than, the quasiparticle linewidth.

The reduction of the third-order nonlinear response to a combination of two-point functions in Eq.~\eqref{eqn:bubble_square_relation} might suggest, at first glance, that third-order spectroscopy offers little additional insight. However, this relation provides a quantitative route to resolving the symmetry of pair continua and identifying their experimental signatures. Momentum-dependent probe form factors enter the linear and nonlinear responses through different combinations, allowing polarization-resolved measurements to separate symmetry components that are not individually resolved in the linear spectrum. For example, products of (p)-wave-like vertices contain (d)-wave-like components that suitable polarization combinations can isolate. This suggests applications to symmetry-resolved spectroscopy of pair excitations in unconventional superconductors and quantum magnets. Relevant magnetic examples include two-magnon continua in ordered antiferromagnets, whose Raman response exhibits polarization-dependent selection rules~\cite{FleuryLoudon1968}, and Majorana-pair continua in Kitaev spin liquids, where the scattering vertices provide symmetry-dependent access to fractionalized excitations~\cite{Perreault2015}. Polarization-selective access to spinon continua has also been demonstrated theoretically in two-dimensional coherent spectroscopy of frustrated magnets~\cite{Potts2024}.

Generalizing the damping to be momentum-dependent but frequency-independent linewidth $\gamma_\mathbf{k}$ preserves the reduction at fixed momentum, although the result can no longer, in general, be expressed solely in terms of the ordinary pair bubble. This case is discussed in~\ref{sec: mom_dep_gamma}. A more fundamental limitation arises from a genuinely frequency-dependent self-energy. When its frequency dependence cannot be absorbed into a quasiparticle pole with renormalized dispersion, residue, and linewidth, the finite-difference identities used in the derivation acquire an explicit dependence on the internal loop frequency and the square no longer closes onto ordinary one-frequency pair bubbles. We discuss this regime in~\ref{sec: freq_sigma}. 

\ref{App: Applications} presents detailed applications of our formalism. We first examine in~\ref{app:interband_pair} echoes arising from interband electronic transitions in semiconductors. We show how the noncommutativity of the Green's functions in Keldysh space entering the third-order response generates an echo despite conservation of the total particle number, and reinterpret this process in terms of electron--hole pair creation. We then recover, in~\ref{sec:tfim} and~\ref{app:twisted_kitaev}, established results for the third-order nonlinear response of the transverse-field Ising model~\cite{Wan2019SpinonEcho} and the twisted Kitaev chain~\cite{Sim2023}, respectively, by directly applying the results of this section.

\subsection{Two-dimensional spectroscopy}
\label{sec:ABB_echo}

% The illustrative spectra in this section use $f_{\mathbf k}=1$, so that $\tilde{\Pi}_0=\Pi_0$.

We now apply the bubble--square relations obtained for free particles in Eq.~\eqref{eqn:bubble_square_relation} to a standard two-dimensional nonlinear spectroscopy protocol. Two excitation pulses, $A$ and $B$, separated by a delay $\tau>0$, interact with the system, which is subsequently allowed to evolve for a detection time $t>0$. The nonlinear signal $S_{\rm nl}(t,\tau)$ is resolved as a function of both time intervals, and a two-dimensional spectrum is obtained by Fourier transforming with respect to $\tau$ and $t$. The corresponding frequencies $\omega_\tau$ and $\omega_t$ resolve, respectively, the coherence generated between the two pulses and the coherence emitted during detection.

At third order, the two-pulse protocol contains contributions proportional to $E_A^2E_B$ and $E_AE_B^2$. We focus here on the latter, conventionally denoted the ABB contribution, in which the system interacts once with pulse $A$ and twice with pulse $B$. Within this sector, one can distinguish rephasing (R) and nonrephasing (NR) pathways according to the relative phase evolution during the two time intervals. In the rephasing sector, the coherences during $\tau$ and $t$ evolve with opposite frequency orientations. Phase accumulated during the first interval can therefore be undone during the second, producing the echo or rephasing peak, analogous to the spin echo~\cite{Hahn_1950}. In the nonrephasing sector, by contrast, the two coherences evolve with the same frequency orientation and no such refocusing occurs~\cite{salvador2024principles, Liu2025Qmat}.

For the ABB sequence, the chronology of the four external vertices is
\begin{equation}
t_1=0,
\qquad
t_2=t_3=\tau,
\qquad
t_4=\tau+t,
\qquad
\tau>0,
\qquad
t>0,
\label{eqn:ABB_times}
\end{equation}
so that the first interaction occurs at pulse $A$, the next two at pulse $B$, and the final vertex corresponds to the emitted signal. The rephasing and nonrephasing contributions correspond to different frequency orientations of these same four vertices, which we consider in turn below.

\subsubsection{Rephasing response and echo ridge}

\begin{figure}
    \centering
    \includegraphics[width=\linewidth]{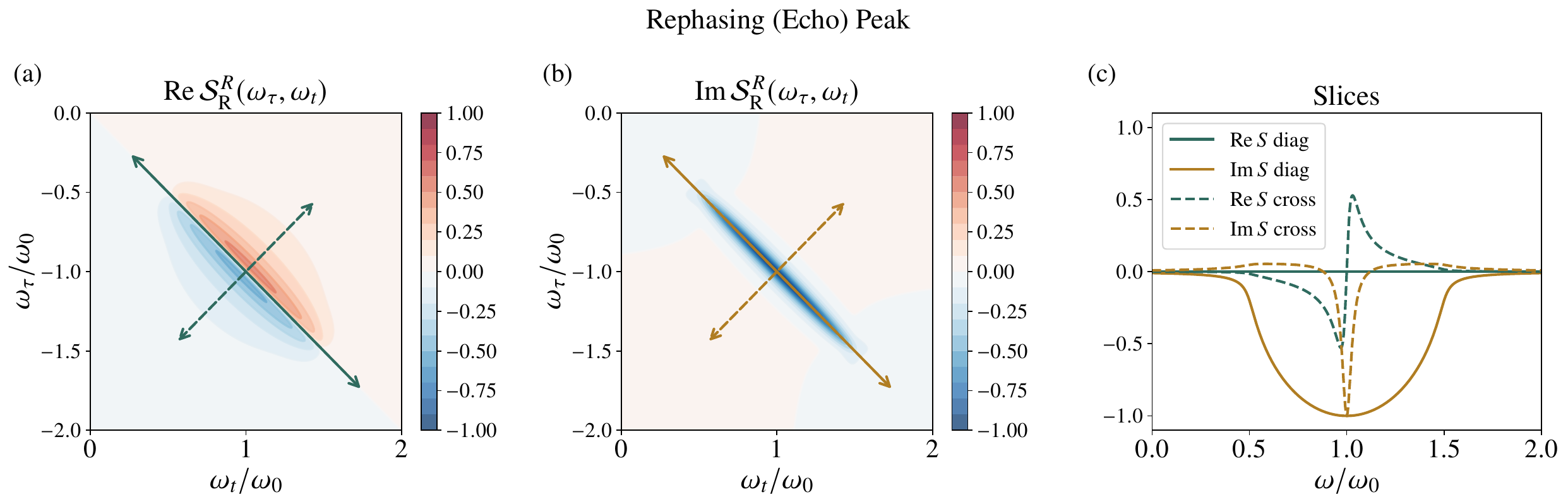}
    \caption{\textbf{Rephasing peak for a free pair continuum.} Normalized real(a) and imaginary (b) parts of the rephasing (echo) peak and slices across the diagonal (solid) and cross-diagonal (dashed) for a toy model with a semicircular density of states. (c) The cut  along the diagonal of the imaginary part corresponds to the spectral function of the weighted pair bubble, while the cross diagonal characterizes the homogeneous damping of the pairs. The echo peak thus allows one to disentangle the continuum from its intrinsic damping, a task inaccessible for traditional linear spectroscopy. In the plots we have used a homogeneous damping of $\gamma/\omega_0=0.015$, and an energy half-bandwidth of the semicircular model of $\delta\omega_0/\omega_0=0.5$, see~\ref{app:semicircular_model} for more details. }
    \label{fig: R_free}
\end{figure}

The rephasing sector of the 2D map can be obtained analytically from Eq.~\eqref{eqn:bubble_square_relation}. The details of the derivation are provided in~\ref{App: R_calculation}. The resulting expression reads
\begin{equation}
\mathcal S_{0,\rm R}^R(\omega_\tau,\omega_t)
=
i\,
\frac{
\tilde{\Pi}_0^R(\omega_t)-\tilde{\Pi}_0^A(-\omega_\tau)
}{
\omega_t+\omega_\tau+4i\gamma
}. 
\label{eqn:ABB_echo_map}
\end{equation}
Equation~\eqref{eqn:ABB_echo_map} makes the rephasing echo explicit. The resonant denominator is minimized when $\omega_t+\omega_\tau=0$, so the signal forms a ridge along the rephasing anti-diagonal $\omega_t=-\omega_\tau$. Along this ridge, writing $\omega_t=-\omega_\tau=\Omega$, gives
\begin{equation}
\mathcal S_{0,\rm R}^R(-\Omega,\Omega)=\frac{1}{4\gamma}\left[\tilde{\Pi}_0^R(\Omega)-\tilde{\Pi}_0^A(\Omega)\right]=-\frac{i\pi}{\gamma}\int_{\mathbf k}|f_{\mathbf k}|^4F_{\eta,\mathbf k}\,\delta_{2\gamma}(\Omega-2\xi_{\mathbf k}).
\label{eqn:ABB_echo_spectral_weight}
\end{equation}
where in the second equality we have employed the weighted counterpart of Eq.~\eqref{eqn:finite_gamma_spectral_jump}. Thus the ABB rephasing trajectory projects the full three-frequency square onto its spectral contribution. The coherence generated by the first pulse is resolved along $\omega_\tau=-\omega_1$, while the emitted coherence is resolved along $\omega_t=\omega_4$. Their refocusing produces the echo ridge $\omega_t=-\omega_\tau$, whose intensity directly follows the spectral weight of the weighted pair bubble and is enhanced as $\gamma^{-1}$. Note the similarities to the case of disordered bosonic excitations presented in Ref.~\cite{salvador2025Echo_theory}.

The linewidth of the echo can be extracted by considering a cut perpendicular to the rephasing ridge. Starting from Eq.~\eqref{eqn:ABB_echo_map}, we consider a point on the echo ridge and parameterize the perpendicular direction by
\begin{equation}
\omega_t=\Omega+\delta,
\qquad
\omega_\tau=-\Omega+\delta.
\end{equation}
Along this cut, the rephasing response becomes
\begin{equation}
\mathcal{S}^{R}_{0,\rm R}(\delta;\Omega)
=
\frac{i}{2}
\frac{
\tilde{\Pi}_0^{R}(\Omega+\delta)
-
\tilde{\Pi}_0^{A}(\Omega-\delta)
}{
\delta+2i\gamma
}.
\end{equation}
Close to the echo ridge, $|\delta|\lesssim 2\gamma$, and provided that the pair spectral weight varies slowly on the scale of the homogeneous linewidth, we may approximate
\begin{equation}
\tilde{\Pi}_0^{R}(\Omega+\delta)
-
\tilde{\Pi}_0^{A}(\Omega-\delta)
\simeq
\tilde{\Pi}_0^{R}(\Omega)-\tilde{\Pi}_0^{A}(\Omega)
=
2i\,\operatorname{Im}\tilde{\Pi}_0^{R}(\Omega),
\end{equation}
and we therefore obtain
\begin{equation}
\mathcal{S}^{R}_{0,R}(\delta;\Omega)
\simeq
-\frac{
\operatorname{Im}\tilde{\Pi}_0^{R}(\Omega)
}{
\delta+2i\gamma
}.
\label{eqn:echo_transverse_lineshape}
\end{equation}
The imaginary part of this response, as well as its squared modulus, has a Lorentzian transverse profile with a half width at half maximum of $2\gamma$ in the coordinate $\delta$, equal to the pair-dephasing rate. This approximation requires the variation of both the spectral and dispersive contributions to the numerator to be negligible, i.e., $2\gamma|\operatorname{Re}\partial_\omega\tilde{\Pi}_0^R(\Omega)|\ll|\operatorname{Im}\tilde{\Pi}_0^R(\Omega)|$. The two directions through the echo peak thus contain complementary information: along the rephasing ridge the response resolves the spectral weight of the weighted pair bubble, whereas the transverse linewidth directly measures the homogeneous pair-dephasing rate. We plot in Fig.~\ref{fig: R_free} the real and imaginary part of the rephasing contribution given in Eq.~\eqref{eqn:ABB_echo_map} and the cuts along the diagonal and cross diagonal corresponding to Eqs.~\eqref{eqn:ABB_echo_spectral_weight} and~\eqref{eqn:echo_transverse_lineshape}.

\subsubsection{Nonrephasing response}
\label{sec:ABB_NR}

\begin{figure}
    \centering
    \includegraphics[width=1\linewidth]{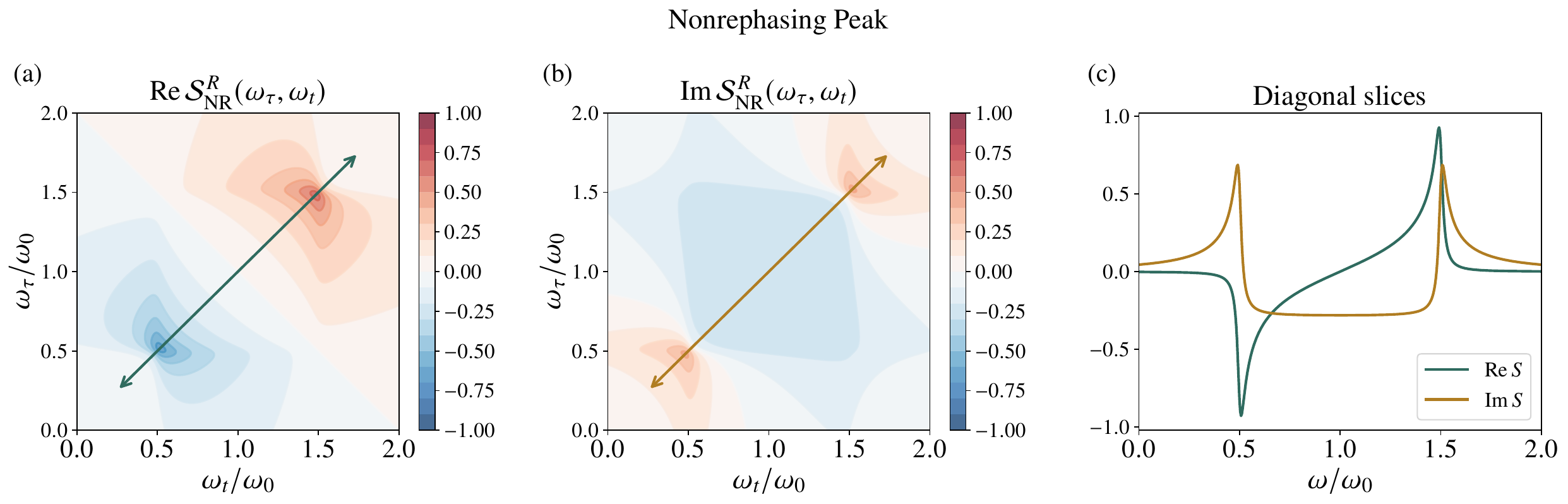}
    \caption{\textbf{Nonrephasing peak for a free pair continuum.} Normalized real (a) and imaginary (b) parts of the nonrephasing peak, together with line cuts (c) along the diagonal. The nonrephasing peak directly probes the frequency derivative of the pair bubble.  In the plots we have used a homogeneous damping of $\gamma/\omega_0=0.015$, and an energy half-bandwidth of the semicircular model of $\delta\omega_0/\omega_0=0.5$, see~\ref{app:semicircular_model} for more details.}
    \label{fig: NR_free}
\end{figure}

The nonrephasing sector of the 2D map can likewise be obtained analytically from Eq.~\eqref{eqn:bubble_square_relation}. The details of the derivation are provided in~\ref{App: NR_calculation}. The resulting expression reads
\begin{equation}
\mathcal S_{0,\rm NR}^R(\omega_\tau,\omega_t)
=
-\frac{i}{2}
\frac{
\tilde{\Pi}_0^R(\omega_t)-\tilde{\Pi}_0^R(\omega_\tau)
}{
\omega_t-\omega_\tau
}.
\label{eqn:NR_ABB_map}
\end{equation}
Equation~\eqref{eqn:NR_ABB_map} should be contrasted with the rephasing result in Eq.~\eqref{eqn:ABB_echo_map}. The nonrephasing response contains only retarded bubbles and is a regular divided difference of $\tilde{\Pi}_0^R$. There is therefore no retarded--advanced spectral discontinuity and no pair of linewidth poles pinching the projected-frequency contour.

The nonrephasing feature lies along the diagonal $\omega_t=\omega_\tau=\omega$. In the coincident-frequency limit, Eq.~\eqref{eqn:NR_ABB_map} becomes
\begin{equation}
\mathcal S_{0,\rm NR}^R(\omega,\omega)
=
-\frac{i}{2}
\partial_\omega\tilde{\Pi}_0^R(\omega).
\label{eqn:NR_ABB_diagonal_response}
\end{equation}
Thus, in contrast with the rephasing echo, the nonrephasing contribution probes the frequency dependence of the retarded pair bubble and does not contain the $\gamma^{-1}$ enhancement associated with the retarded--advanced pinch. We plot in Fig.~\ref{fig: NR_free} the real and imaginary part of the nonrephasing contribution given in Eq.~\eqref{eqn:NR_ABB_map} and the cut along the $\omega_\tau=\omega_t=\omega$ diagonal obtained in Eq.~\eqref{eqn:NR_ABB_diagonal_response}.

\section{Particle--particle interactions and collective modes}
\label{sec:particle_particle_interactions}

We now extend the preceding analysis of free bosonic and fermionic quasiparticles to include interactions within the same particle–particle channel. We consider a separable quartic interaction of the form
\begin{equation}
H_{\mathrm{int}}=g\int_{\mathbf k,\mathbf k'}u_{\mathbf k}^*u_{\mathbf k'}P_{\mathbf k}^\dagger P_{\mathbf k'},
\end{equation}
with $g>0$ ($g<0$) corresponding to repulsion (attraction). Here $P_{\mathbf k}=\frac12\psi_{\mathbf k}^T\Lambda\psi_{-\mathbf k}$ is the momentum-resolved pair bilinear introduced in Sec.~\ref{sec:bubble_square_relation}, while $u_{\mathbf k}$ specifies the interaction channel. The interaction form factor is independent of the drive form factor $f_{\mathbf k}$, which enters only through the external source. 
We assume $u_{-\mathbf k}=u_{\mathbf k}$ and recover the contact interaction for $u_{\mathbf k}=1$. More generally, nonlocal pairing interactions can be projected onto separable symmetry channels. For instance, nearest-neighbor interactions on a square lattice naturally give rise to extended $s$-wave and $d_{x^2-y^2}$ channels, characterized by $u_{\mathbf k}=\cos k_x+\cos k_y$ and $u_{\mathbf k}=\cos k_x-\cos k_y$, respectively. Such symmetry channels arise naturally in square-lattice models of strongly correlated electrons, particularly in descriptions of unconventional superconductivity driven by antiferromagnetic exchange~\cite{KotliarLiu1988,LeeNagaosaWen2006,Scalapino2012}, while related exchange physics also governs magnetic correlations and bimagnon excitations in doped Mott insulators relevant to the cuprates~\cite{andrei2026universalmagneticenergyscale}. Retaining a single symmetry channel provides a simple separable approximation to the underlying momentum-dependent interaction, although the formalism can be generalized to include multiple coupled symmetry channels.

Furthermore, we retain repeated particle–particle ladder insertions within the random phase approximation (RPA), keeping the single-particle propagators and their linewidth $\gamma$ fixed and neglecting self-energy and crossed-channel vertex corrections. Within this approximation, the interaction dresses both the linear pair susceptibility and the external vertices of the nonlinear square. We first derive these two contributions and show that the bubble–square relation survives the resummation, yielding Eq.~\eqref{eqn:interacting_bubble_square_relation}. We then examine the emergence of an isolated bound state and show that, in the strong-binding regime, the bare square generates an effective local quartic interaction for the collective field, Eq.~\eqref{eqn:phi4_tree}. Finally, we derive the corresponding rephasing and nonrephasing spectra and compare them with the microscopic response, tracing how the transfer of spectral weight into the bound state transforms the extended continuum echo ridge into a localized collective-mode feature.

\subsection{Particle--particle ladder}

\begin{figure}
    \centering
    \includegraphics[width=0.95\linewidth]{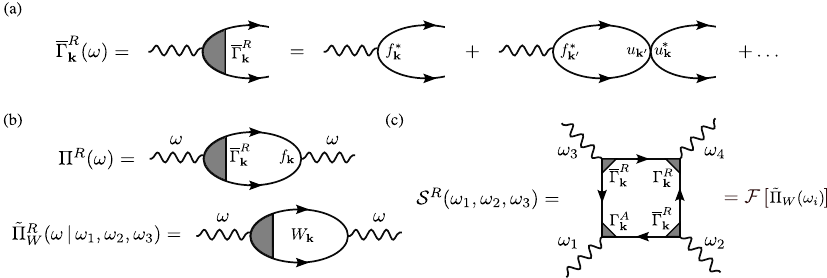}
    \caption{ \textbf{RPA diagrammatics for the pair-bubble and square.} (a) Ladder correction to the pair creation vertex; the annihilation vertex is defined analogously. (b) Dressed particle-particle bubble and (c) square diagram with vertex corrections.}
    \label{fig: diagrams_RPA}
\end{figure}

Repeated scattering in the particle--particle channel, cf. Fig.~\ref{fig: diagrams_RPA}(b), generates the dressed pair-bubble. To keep track of the drive and interaction vertices, we define
\begin{equation}
\Pi_{ab,0}^{R/A}(\omega)=2\int_{\mathbf k}a_{\mathbf k}^*b_{\mathbf k}\frac{F_{\eta,\mathbf k}}{\omega-2\xi_{\mathbf k}\pm2i\gamma},\qquad a,b\in\{u,f\},\qquad \Pi_{ff,0}^{R/A}=\Pi_0^{R/A}.
\label{eqn:mixed_pair_bubbles}
\end{equation}
Each bubble contains the form factors of its two attached vertices: $f$ for a drive vertex and $u$ for an interaction vertex. The separable kernel $V_{\mathbf k\mathbf k'}=g u_{\mathbf k}^*u_{\mathbf k'}$ keeps the ladder resummation scalar. The corresponding linear susceptibility is
\begin{equation}
\Pi^{R/A}(\omega)
=\Pi_0^{R/A}(\omega)+g\Pi_{fu,0}^{R/A}(\omega)\mathcal L^{R/A}(\omega)\Pi_{uf,0}^{R/A}(\omega), \qquad
\mathcal L^{R/A}(\omega)=\frac{1}{1-g\Pi_{uu,0}^{R/A}(\omega)}.
\label{eqn:resummed_pair_bubble}
\end{equation}
Here $\mathcal L^{R/A}$ is the vertex-renormalization factor, see Fig.~\ref{fig: diagrams_RPA}. For real $g$, $\mathcal L^A(\omega)=[\mathcal L^R(\omega)]^*$ and $\Pi^A(\omega)=[\Pi^R(\omega)]^*$. For $f_{\mathbf k}=u_{\mathbf k}$, all the bubbles in Eq.~\eqref{eqn:mixed_pair_bubbles} coincide and we recover the familiar form for the dressed susceptibility $\Pi^{R/A}(\omega)=\Pi_0^{R/A}(\omega)/[1-g\Pi_0^{R/A}(\omega)]=\mathcal L^{R/A}(\omega)\Pi_0^{R/A}(\omega)$. In particular, $u_{\mathbf k}=f_{\mathbf k}=1$ recovers the standard contact-interaction scalar ladder.

\subsection{Vertex resummation of the square diagram}

We next consider the effect of the particle--particle interaction on the four-propagator square. Within the present ladder approximation, the constituent square derived in Eq.~\eqref{eqn:bubble_square_relation} plays the role of the irreducible nonlinear vertex. Repeated particle--particle scattering dresses each external probe vertex according to
\begin{equation}
\begin{aligned}
\Gamma_{\mathbf k}^{\alpha}(\omega)&=f_{\mathbf k}+u_{\mathbf k}g\mathcal L^{\alpha}(\omega)\Pi_{uf,0}^{\alpha}(\omega),\\
\bar\Gamma_{\mathbf k}^{\alpha}(\omega)&=f_{\mathbf k}^*+u_{\mathbf k}^*g\mathcal L^{\alpha}(\omega)\Pi_{fu,0}^{\alpha}(\omega),\qquad \alpha\in\{R,A\}, 
\end{aligned}
\label{eqn:dressed_probe_vertices}
\end{equation}
see Fig.~\ref{fig: diagrams_RPA}. Here $\Gamma_{\mathbf k}^{\alpha}$ and $\bar\Gamma_{\mathbf k}^{\alpha}$ denote the annihilation and creation vertices, respectively, with $\bar\Gamma_{\mathbf k}^{A}(\omega)=[\Gamma_{\mathbf k}^{R}(\omega)]^*$. 
% For $f_{\mathbf k}=u_{\mathbf k}$, the vertices reduce to $\Gamma_{\mathbf k}^\alpha=u_{\mathbf k}\mathcal L^\alpha$ and $\bar\Gamma_{\mathbf k}^\alpha=u_{\mathbf k}^*\mathcal L^\alpha$; for a general drive form factor the correction is additive and cannot be factored out of the momentum integral.
For the ordered rephasing contribution considered in the previous section, legs $1$ and $4$ are outgoing while legs $2$ and $3$ are incoming, with $\omega_1+\omega_4=\omega_2+\omega_3$, cf. Eq.~\eqref{eqn: frequency_conservation}. The causal orientation of the square derived in Sec.~\ref{sec:bubble_square_relation} gives an advanced pair vertex associated with $\omega_1$ and retarded pair vertices associated with $\omega_2$, $\omega_3$, and $\omega_4$. The vertex-resummed square is therefore, cf. Fig.~\ref{fig: diagrams_RPA}(c),
\begin{equation}
\mathcal S^R(\omega_1,\omega_2,\omega_3)=\int_{\mathbf k}W_{\mathbf k}(\omega_1,\omega_2,\omega_3)\,s_{0,\mathbf k}^R(\omega_1,\omega_2,\omega_3),\quad \text{with} \quad
W_{\mathbf k}(\omega_1,\omega_2,\omega_3)=\Gamma_{\mathbf k}^{A}(\omega_1)\bar\Gamma_{\mathbf k}^{R}(\omega_2)\bar\Gamma_{\mathbf k}^{R}(\omega_3)\Gamma_{\mathbf k}^{R}(\omega_4),
\label{eqn:resummed_square_factorized}
\end{equation}
where $s_{0,\mathbf k}^R$ is the fixed-momentum constituent kernel before attaching the external vertices, so that $\mathcal S_0^R=\int_{\mathbf k}|f_{\mathbf k}|^4s_{0,\mathbf k}^R$, see Eq.~\eqref{eqn:bubble_square_relation}. Since the dressed vertices are independent of the internal loop frequency, the reduction at fixed momentum remains unchanged. It is useful to define
\begin{equation}
\tilde{\Pi}_{W}^{R/A}(\omega\mid\omega_1,\omega_2,\omega_3)=2\int_{\mathbf k}W_{\mathbf k}(\omega_1,\omega_2,\omega_3)\frac{F_{\eta,\mathbf k}}{\omega-2\xi_{\mathbf k}\pm2i\gamma},
\label{eqn:dressed_weight_bubble}
\end{equation}
where the weight $W_{\mathbf k}$ contains all four dressed drive vertices and reduces to $|f_{\mathbf k}|^4$ for $g=0$. Defining, as before, $\Delta_{31}\equiv\omega_3-\omega_1$ and $\Delta_{21}\equiv\omega_2-\omega_1$, the complete interacting bubble--square relation can be written as
\begin{equation}
\mathcal S^R(\omega_1,\omega_2,\omega_3)=\frac{\tilde{\Pi}_{W}^R(\omega_4)-\tilde{\Pi}_{W}^R(\omega_2)}{\Delta_{31}(\Delta_{21}+2i\gamma)}
+\frac{\tilde{\Pi}_{W}^R(\omega_4)-\tilde{\Pi}_{W}^R(\omega_3)}{(\Delta_{31}+2i\gamma)\Delta_{21}}+\frac{\tilde{\Pi}_{W}^A(\omega_1)-\tilde{\Pi}_{W}^R(\omega_4)}{(\Delta_{31}+2i\gamma)(\Delta_{21}+2i\gamma)},
\label{eqn:interacting_bubble_square_relation}
\end{equation}
where the common arguments $(\omega_1,\omega_2,\omega_3)$ of each weighted bubble $\tilde{\Pi}_{W}^R$ have been suppressed. Eq.~\eqref{eqn:interacting_bubble_square_relation} is the RPA interacting generalization of Eq.~\eqref{eqn:bubble_square_relation}, which is directly recovered for $g\rightarrow0$, when $W_{\mathbf k}=|f_{\mathbf k}|^4$. In particular, it shows that the reduction to two-point functions survives the interaction resummation. For a specified interaction strength and quasiparticle linewidth, the response is determined by the mixed pair bubbles entering the dressed probe vertices in Eq.~\eqref{eqn:mixed_pair_bubbles} and the four-vertex-weighte bubbles in Eq.~\eqref{eqn:dressed_weight_bubble}, without an independent calculation of any irreducible four-point vertex. In the special case of $f_{\mathbf k}=u_{\mathbf k}=1$, all the mixed pair bubbles in Eq.~\eqref{eqn:mixed_pair_bubbles} coincide, $\Pi_{ff,0}^{\alpha}=\Pi_{fu,0}^{\alpha}=\Pi_{uf,0}^{\alpha}=\Pi_{uu,0}^{\alpha}\equiv\Pi_0^{\alpha}$, the dressed vertices in Eq.~\eqref{eqn:dressed_probe_vertices} therefore become momentum independent, $\Gamma_{\mathbf k}^{\alpha}(\omega)=\bar\Gamma_{\mathbf k}^{\alpha}(\omega)=1+g\mathcal L^{\alpha}(\omega)\Pi_0^{\alpha}(\omega)=\mathcal L^{\alpha}(\omega)$, and the four-vertex weight in Eq.~\eqref{eqn:resummed_square_factorized} no longer depends on $\mathbf k$ and can be taken outside the momentum integral,
$W_{\mathbf k}(\omega_1,\omega_2,\omega_3)\equiv W(\omega_1,\omega_2,\omega_3)=\mathcal L^A(\omega_1)\mathcal L^R(\omega_2)\mathcal L^R(\omega_3)\mathcal L^R(\omega_4).
% \label{eqn:contact_limit_weight}
$
Substituting into Eq.~\eqref{eqn:interacting_bubble_square_relation} gives
\begin{equation}
\mathcal S^R(\omega_1,\omega_2,\omega_3)=\mathcal L^A(\omega_1)\mathcal L^R(\omega_2)\mathcal L^R(\omega_3)\mathcal L^R(\omega_4)\,\mathcal S_0^R(\omega_1,\omega_2,\omega_3).
\label{eqn:contact_limit_square_factorization}
\end{equation}

% As we will see next, the emergence of a collective pole due to the interactions affects several frequency arguments of the nonlinear response, reshaping the multidimensional spectrum. This provides a direct connection between the bound-state formation visible in linear spectroscopy and its nonlinear signatures, while establishing a reference against which additional correlations beyond the present RPA treatment can be identified.

\begin{figure}
    \centering
    \includegraphics[width=1\linewidth]{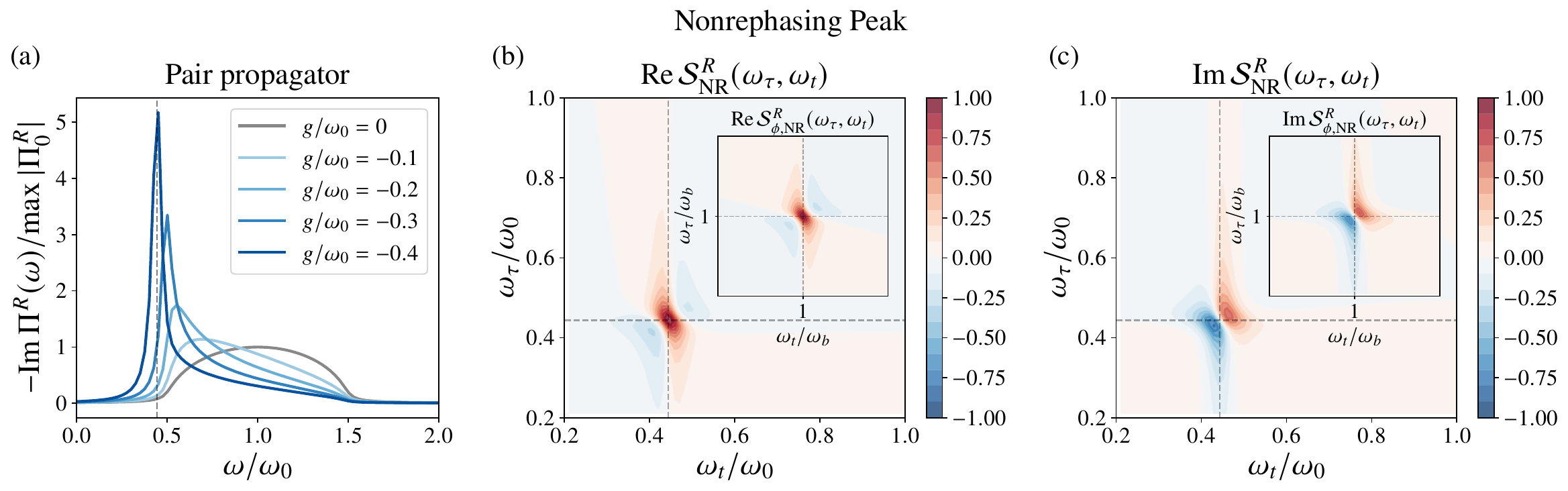}
    \caption{(a) Pair spectral function as a function of interaction strength, showing the evolution from the noninteracting continuum to a bound-state resonance with increasing attraction. (b,c) Real and imaginary parts, respectively, of the nonrephasing two-dimensional spectrum obtained numerically from the interacting response in Eq.~\eqref{eqn:interacting_bubble_square_relation}, for $g/\omega_0=-0.4$. The insets show the analytical collective-mode spectra from Eq.~\eqref{eqn:phi4_NR_map}. In the plots we have used a homogeneous damping of $\gamma/\omega_0=0.015$, and an energy half-bandwidth of the semicircular model of $\delta\omega_0/\omega_0=0.5$, see~\ref{app:semicircular_model} for more details.}
    \label{fig: NR_interacting}
\end{figure}

\subsubsection{Strong-binding limit and effective collective-mode description}
\label{sec:strong_binding_phi4}

For sufficiently strong attraction, the ladder develops an isolated bound-state pole below the two-particle continuum. In this limit, it is convenient to introduce the collective-mode propagator
\begin{equation}
\mathcal{D}^{R/A}(\omega)\equiv g\mathcal{L}^{R/A}(\omega)=\frac{g}{1-g\Pi_{uu,0}^{R/A}(\omega)}.
\label{eqn:collective_D}
\end{equation}
Close to an isolated collective mode at frequency $\omega_b$, this propagator takes the pole form
\begin{equation}
\mathcal{D}^{R/A}(\omega)\simeq\frac{Z_{\mathcal{D}}}{\omega-\omega_b\pm i\Gamma_b},
\label{eqn:D_pole}
\end{equation}
where $Z_{\mathcal{D}}$ is the collective-mode residue and $\Gamma_b=2\gamma$ its homogeneous linewidth. The probe couples to this mode through the mixed bubbles in Eq.~\eqref{eqn:dressed_probe_vertices}. Writing $z_b=\omega_b-i\Gamma_b$ and defining
\begin{equation}
c_f\equiv g\Pi_{uf,0}^{R}(z_b),
\label{eqn:collective_probe_overlap}
\end{equation}
the pole condition gives $c_f=1$ for $f_{\mathbf k}=u_{\mathbf k}$. Because $\Gamma_b=2\gamma$, the denominators of the mixed bubbles evaluated at $z_b$ are real below the continuum, and $g\Pi_{fu,0}^{R}(z_b)=c_f^*$. Near the isolated pole, the singular parts of the probe vertices are therefore $\Gamma_{\mathbf k}^{R/A}\simeq u_{\mathbf k}c_f\mathcal D^{R/A}/g$ and $\bar\Gamma_{\mathbf k}^{R/A}\simeq u_{\mathbf k}^*c_f^*\mathcal D^{R/A}/g$. Retaining the contribution in which all four external vertices are resonant gives
\begin{equation}
\mathcal S^R(\omega_1,\omega_2,\omega_3)\simeq{}\frac{|c_f|^4\mathcal S_{u,0}^{R}(\omega_1,\omega_2,\omega_3)}{g^4}\mathcal D^{A}(\omega_1)\mathcal D^{R}(\omega_2)\mathcal D^{R}(\omega_3)\mathcal D^{R}(\omega_4),
\label{eqn:dressed_square_D}
\end{equation}
where $\mathcal S_{u,0}^R\equiv\int_{\mathbf k}|u_{\mathbf k}|^4s_{0,\mathbf k}^R$ is the constituent square with four interaction-channel vertices. Note that the four drive overlaps are included in $|c_f|^4$. For a general drive form factor, terms with fewer resonant vertices provide additional contributions to Eq.~\eqref{eqn:resummed_square_factorized}; for $f_{\mathbf k}=u_{\mathbf k}$, the four-vertex factorization in Eq.~\eqref{eqn:dressed_square_D} is exact with $c_f=1$, even before making the pole approximation.

\begin{figure}[t]
    \centering
    \includegraphics[width=\linewidth]{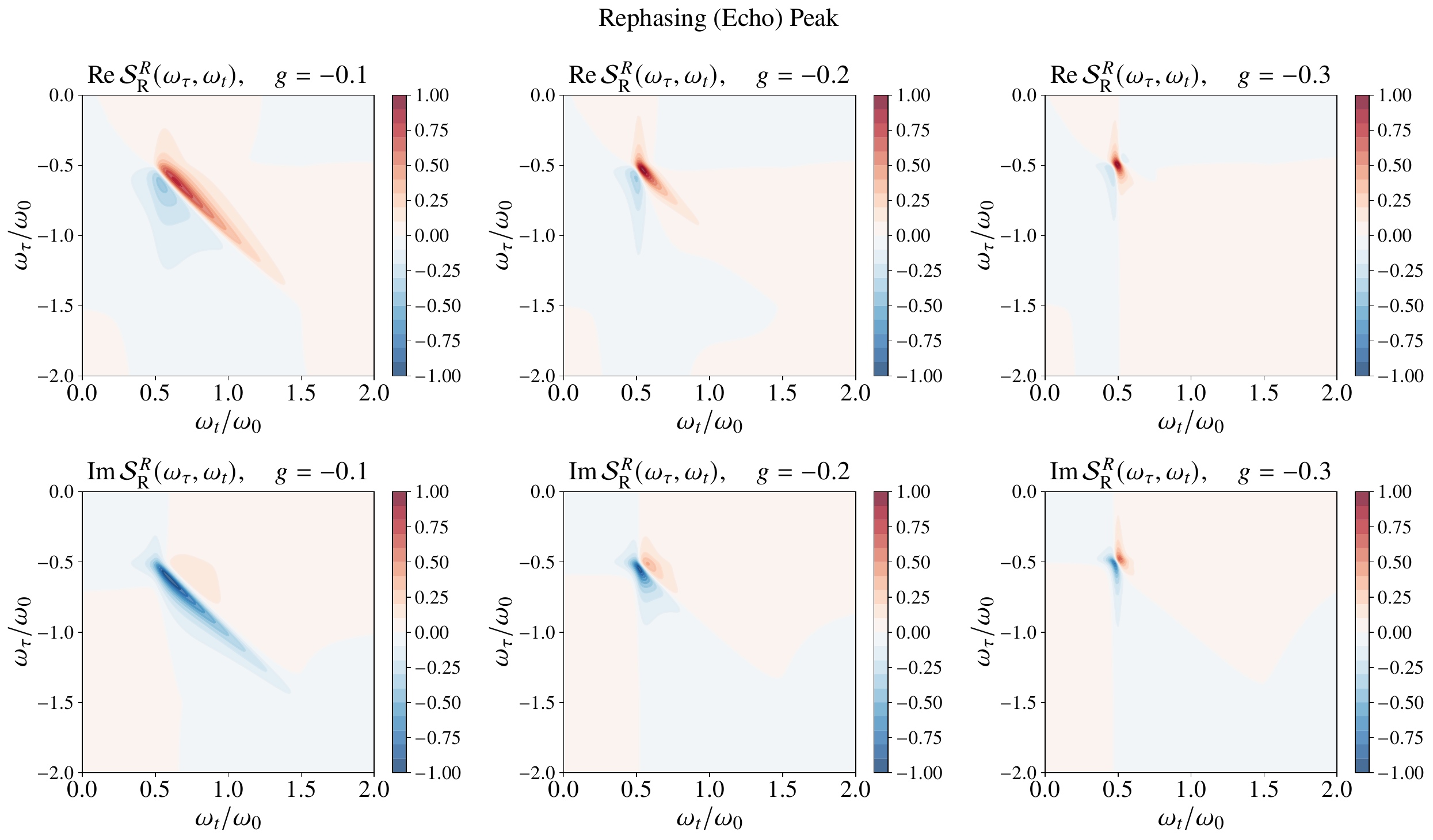}
    \caption{\textbf{Signatures of the appearance of the collective mode in the echo ridge.} Evolution of the real (top) and imaginary (bottom) parts of the echo peak as the interaction strength is increased. As the attraction grows the collective mode starts distorting the echo peak ridge from a purely rephasing signature, cf. Fig.~\ref{fig: R_free}, to a collective mode, cf. Fig.~\ref{fig: NR_interacting}. In the imaginary part, positive regions develop around the echo, signaling the emergence of the collective-mode pole in the interacting pair susceptibility $\Pi^R(\omega)$. In the plots we have used a homogeneous damping of $\gamma/\omega_0=0.015$, and and energy half-bandwidth of the semicircular model of $\delta\omega_0/\omega_0=0.5$, see~\ref{app:semicircular_model} for more details.}
    \label{fig: interacting_R}
\end{figure}

When the bound state is well separated from the constituent continuum, the four collective propagators restrict the relevant external frequencies to a narrow region around $\omega_b$. If the irreducible constituent square is smooth on this frequency scale, it may be replaced by its value at the collective pole $\lambda_{\phi}\simeq\mathcal S_{u,0}^{R}(\omega_b,\omega_b,\omega_b)/g^4$, where $\lambda_{\phi}$ defines the effective quartic interaction between the emergent collective modes. For nonzero probe overlap $c_f$, the leading resonant nonlinear response then reduces to
\begin{equation}
\mathcal S^R(\omega_1,\omega_2,\omega_3)\simeq |c_f|^4\lambda_{\phi}\mathcal D^{A}(\omega_1)\mathcal D^{R}(\omega_2)\mathcal D^{R}(\omega_3)\mathcal D^{R}(\omega_4).
\label{eqn:phi4_tree}
\end{equation}
Equation~\eqref{eqn:phi4_tree} is precisely the tree-level four-point response of an effective interacting collective field with a local $|\phi|^4$ vertex and four probe-overlap factors. If $c_f=0$, the interaction-channel collective mode has no pole residue in this probe channel. We provide in~\ref{app:semicircular_model} a detailed derivation of boundstate energy $\omega_b$ and quasiparticle weight $Z_b$ for the contact-interaction semicircular toy model used in Figs.~\ref{fig: NR_interacting} and~\ref{fig: interacting_R}.

\subsection{Two-dimensional spectroscopy}

We now investigate how the formation of a collective mode reshapes the two-dimensional response. As the attraction increases, spectral weight is transferred from the pair continuum to an emerging collective mode, modifying both the rephasing and nonrephasing spectra. For general interaction strengths, we evaluate the maps numerically using Eqs.~\eqref{eqn:ABB_kernel} and~\eqref{eqn:NR_ABB_time_kernel}, together with Eq.~\eqref{eqn:interacting_bubble_square_relation}. The figures are obtained in the contact-interaction limit $u_{\mathbf k}=f_{\mathbf k}=1$, while the separable interaction with general $u_{\mathbf k}$ can be evaluated using the same expressions. The collective-mode description introduced above provides analytical expressions against which these numerical results can be compared.

Within the pole approximation of Eq.~\eqref{eqn:D_pole} and the local-vertex approximation of Eq.~\eqref{eqn:phi4_tree}, the nonrephasing contribution takes the form
\begin{equation}
\mathcal{S}^{R}_{\phi,{\rm NR}}(\omega_\tau,\omega_t) =
\frac{|c_f|^4\lambda_\phi Z_{\mathcal{D}}^4}{2\Gamma_b} \frac{1}{\omega_\tau-\omega_b+i\Gamma_b} \left[ \frac{1}{\omega_t-\omega_b+i\Gamma_b}-\frac{1}{\omega_t-\omega_b+3i\Gamma_b}\right],
\label{eqn:phi4_NR_map}
\end{equation}
up to an overall convention-dependent numerical factor associated with the definition of the quartic vertex. Under the same approximations, performing the rephasing projection and the one-sided Fourier transforms gives
\begin{equation}
\mathcal{S}^{R}_{\phi,{\rm R}}(\omega_\tau,\omega_t) =\frac{|c_f|^4\lambda_\phi Z_{\mathcal{D}}^4}{2\Gamma_b}\frac{1}{\omega_\tau+\omega_b+i\Gamma_b}\left[\frac{1}{\omega_t-\omega_b+i\Gamma_b}-\frac{1}{\omega_t-\omega_b+3i\Gamma_b}\right],
\label{eqn:phi4_R_map}
\end{equation}
with the same overall normalization convention as Eq.~\eqref{eqn:phi4_NR_map}.

The two expressions have identical detection-frequency dependence, while their first-delay resonances occur at opposite frequencies. The collective features are therefore centered near $(\omega_\tau,\omega_t)=(\omega_b,\omega_b)$ in the nonrephasing sector and $(-\omega_b,\omega_b)$ in the rephasing sector, as expected.

Fig.~\ref{fig: NR_interacting}(a) illustrates the emergence of a bound-state resonance below the pair continuum in the linear spectral function $-\operatorname{Im}\Pi^R(\omega)$. The corresponding nonrephasing spectrum is shown in Figs.~\ref{fig: NR_interacting}(b,c), together with the analytical collective-mode maps in the insets. For the parameters shown, the dominant nonlinear feature is concentrated near the bound-state resonance, and the analytical expressions, in the insets of Fig.~\ref{fig: NR_interacting}, display the qualitative agreement with a single isolated pole. This comparison connects the redistribution of spectral weight in the linear response to the emergence of a localized collective feature in the two-dimensional spectrum, while testing the pole and local-vertex approximations underlying the effective description.

Fig.~\ref{fig: interacting_R} shows how the rephasing spectrum evolves with increasing attraction. At weaker coupling, the response retains an extended ridge along $\omega_t=-\omega_\tau$, reflecting the underlying pair continuum. As the attraction increases, the signal becomes concentrated near the lower continuum edge and, following bound-state formation, near the collective resonance at $(-\omega_b,\omega_b)$. The real and imaginary parts develop a more localized, sign-changing structure, illustrating how interactions modify both the spectral weight and the phase of the nonlinear response. In particular, positive regions appear around the predominantly negative imaginary echo ridge. This evolution connects the continuum rephasing signal to the localized collective-mode response described by Eq.~\eqref{eqn:phi4_R_map}. The rephasing frequency assignment is preserved, while the extended ridge progressively gives way to a feature dominated by the bound-state resonance.

%%%%%%%%%%%%%%%%%%%%%%%%%%%%%%%%%%%%%%%%%%%%%%%%%%%%%%%%%%%%%%%%%%%%%%%%%%%%%%%%%%%%%%

\section{Conclusions and Outlook}

In this work we have shown that, within the quasiparticle approximation, the third-order response of a pair continuum can be expressed in terms of generalized two-point correlation functions. Specifically, the four-propagator square reduces to a combination of retarded and advanced pair bubbles. This bubble--square relation applies to both bosonic and fermionic quasiparticles and follows from the pole form of the single-particle propagators. For momentum-independent pair vertices and linewidths, the generalized bubbles coincide with those entering the physical linear susceptibility, up to known vertex prefactors.

More generally, the reduction does not make the nonlinear measurement equivalent to linear response. Momentum-dependent probe vertices enter the two signals through different weights: $|f_k|^2$ in the linear pair bubble and $|f_k|^4$ in the echo. Different pulse polarizations generate further combinations of vertices, providing a route to separating momentum and symmetry components that are not individually resolved in the linear spectrum. The nonlinear response can therefore reveal additional information about the pair continuum while remaining expressible through two-point functions. Our relation identifies these functions and provides a quantitative basis for interpreting polarization-resolved echoes.

Applied to two-pulse spectroscopy, the construction separates the information carried by the rephasing and nonrephasing pathways. The rephasing response contains a retarded--advanced spectral contribution and forms an echo ridge whose intensity follows the spectral function of the generalized pair bubble. When the spectral envelope varies slowly on the scale of the homogeneous linewidth, the transverse profile measures the pair-dephasing rate. The nonrephasing response instead contains only retarded bubbles, highlighting the absence of an echo, but providing access to their frequency dependence. 
% These pathways thus separate the distribution of pair energies from the decay of pair coherence, which are intertwined in the linear spectrum.

The applications presented in~\ref{App: Applications} further illustrate the scope of this construction. Interband electronic transitions can be recast in terms of electron--hole pair creation, while the transverse-field Ising chain provides an exactly solvable example in which the spinon echo is recovered from generalized pair bubbles. The twisted Kitaev chain extends this analysis to multiple quasiparticle bands, demonstrating that mixed pair-creation and interband-transfer pathways, including off-diagonal spectral features, can also be expressed in terms of weighted two-point functions.

We then included particle--particle interactions within a ladder approximation. The reduction survives because the dressed probe vertices are independent of the internal loop frequency and can themselves be expressed through mixed pair bubbles involving the drive and interaction form factors. For attraction, this framework describes the transfer of spectral weight from an extended continuum into an isolated bound state. In the strong-binding regime, the response near the isolated pole reduces to that of an effective interacting collective field. Its propagator follows from the dressed pair susceptibility, while its local quartic interaction is generated by the constituent square. This connects the continuum echo to the nonlinear response of a collective bosonic mode.

We have also shown that momentum-dependent linewidths preserve the pole form of the propagators and hence the reduction at fixed momentum. The response then involves momentum-resolved pair bubbles, and the measured echo can contain a superposition of homogeneous decay rates. A frequency-dependent self-energy presents a different limitation since the coefficients in the resolvent identities acquire an internal-frequency dependence, so the square no longer generally closes onto ordinary pair bubbles. An approximate reduction remains possible when the spectral function is dominated by a well-defined quasiparticle pole and the self-energy can be linearized over the relevant frequency range, yielding renormalized dispersions, residues, and linewidths. The appendices analyze these extensions and their domains of validity.

Several directions emanating from this work remain open. A concrete next step is to construct polarization combinations that isolate symmetry components of the generalized pair response, and determine which provide information not fixed by linear measurements. The finite-difference structure also motivates a search for sum rules relating convergent frequency moments or integrals of the nonlinear response to weighted pair spectral functions. A natural direction for future work is to apply the interacting framework to experimentally relevant pair continua, including excitonic bound states in semiconductors and spinon- or magnon-pair excitations in quantum magnets. Extending the present ladder approximation to multiple coupled symmetry channels and more general interaction kernels would further clarify the extent to which the bubble--square reduction survives beyond the approximations considered here. Additionally, the present formalism could be extended to probes involving higher-order quasiparticle operators, such as transverse-spin spectroscopy of the XXZ chain, where the nonlocal Jordan--Wigner string generates fermionic operators of arbitrarily high order \cite{Li_2021}.

Extending the framework to anyon--anti-anyon pairs could connect it to complementary spectroscopic probes of fractional statistics. Whereas linear-response threshold behavior can encode anyonic statistics~\cite{Morampudi_2017}, nonlinear spectroscopy can access braiding processes involving excitations created by different pulses~\cite{mcginley2022signatures,kirchner2025measuringanyonicexchangephases}. An interesting direction is to establish how these processes enter the rephasing response and its two-dimensional spectrum within the present framework.
More broadly, once the probe vertices and single-particle spectral properties are independently constrained, the present results provide a benchmark for investigating deviations associated with vertex corrections and correlations beyond the quasiparticle and ladder approximations.

\section*{Acknowledgements}

We thank R. Andrei, A. Cavalleri, D. de Santis, F. Essler, A. Liu, F. Marijanovic, I. Morera, S. Parameswaran, and D. Pavicevic for insightful discussions. We acknowledge financial support from the Swiss National Science Foundation (SNSF) through Project No.~200021\_212899, Sinergia Grant No.~CRSII--222792, and the National Centre of Competence in Research (NCCR) Precision (Grant No.~51AU-0\_229299), as well as from the Swiss State Secretariat for Education, Research and Innovation (SERI) under Contract No.~UeM019-1. A.G.S. acknowledges partial support from the Engineering and Physical Sciences Research Council (EPSRC) under Grant No. EP/X030881/1.

\newpage

\appendix

\section{Calculation of the 2D maps}

In this Appendix, we present the detailed derivation of the ABB pathway 2D maps for noninteracting particles, as discussed in Sec.~\ref{sec:bubble_square_relation}. We first consider the rephasing sector, followed by an analogous calculation for the nonrephasing sector, recovering the analytical expressions given in Eqs.~\eqref{eqn:ABB_echo_map} and~\eqref{eqn:NR_ABB_map}, respectively.

\subsection{Rephasing response and echo ridge}
\label{App: R_calculation}

For the rephasing ordering considered in the main text, legs $1$ and $4$ are outgoing while legs $2$ and $3$ are incoming. The external Fourier phase is therefore
\begin{equation}
e^{-i\omega_1t_1}
e^{+i\omega_2t_2}
e^{+i\omega_3t_3}
e^{-i\omega_4t_4}
=
e^{+i\omega_1\tau}
e^{-i\omega_4t}.
\label{eqn:ABB_phase}
\end{equation}
The first delay $\tau$ therefore carries the coherence generated by the first pulse, while the second time $t$ carries the emitted frequency $\omega_4$. The ABB rephasing (R) signal can consequently be written as
\begin{equation}
\mathcal {S}_{0,\rm R}^R(t,\tau)
=
\int\frac{d\omega_1}{2\pi}
\frac{d\omega_2}{2\pi}
\frac{d\omega_3}{2\pi}\,
e^{+i\omega_1\tau}
e^{-i\omega_4t}
\mathcal S_0^R(\omega_1,\omega_2,\omega_3).
\label{eqn:ABB_time_response}
\end{equation}
Importantly, after imposing the ABB trajectory the Fourier phase depends only on $\omega_1$ and $\omega_4$. One independent combination of $\omega_2$ and $\omega_3$ is therefore not resolved and must be integrated over. To make this projection explicit, we introduce
\begin{equation}
\delta\equiv\omega_2-\omega_1, \qquad \text{such that} \qquad
\omega_2=\omega_1+\delta,
\qquad
\omega_3=\omega_4-\delta,
\label{eqn:ABB_frequency_routing}
\end{equation}
and the transformation $(\omega_1,\omega_2,\omega_3)\rightarrow(\omega_1,\omega_4,\delta)$ has unit Jacobian in magnitude. Equation~\eqref{eqn:ABB_time_response} therefore becomes
\begin{equation}
\mathcal S_{0,\rm R}^R(t,\tau)
=
\int\frac{d\omega_1}{2\pi}
\frac{d\omega_4}{2\pi}\,
e^{+i\omega_1\tau}
e^{-i\omega_4t}
K_{\rm R}^R(\omega_1,\omega_4),
\quad 
\text{where}
\quad
K_{\rm R}^R(\omega_1,\omega_4)
\equiv
\int\frac{d\delta}{2\pi}\,
\mathcal S_0^R
\left(
\omega_1,
\omega_1+\delta,
\omega_4-\delta
\right).
\label{eqn:ABB_kernel}
\end{equation}
Thus the restriction $t_2=t_3$ in the time domain becomes a projection in frequency space. We now evaluate Eq.~\eqref{eqn:ABB_kernel} using Eq.~\eqref{eqn:bubble_square_relation}. Defining $s\equiv\omega_4-\omega_1$, the projected kernel separates into
\begin{equation}
K_{\rm R}^R(\omega_1,\omega_4)
=
I_1+I_2+I_{\mathrm{sp}},
\label{eqn:ABB_kernel_decomposition}
\end{equation}
where
\begin{equation}
I_1
=
\int\frac{d\delta}{2\pi}
\frac{
\tilde{\Pi}_0^R(\omega_4)-\tilde{\Pi}_0^R(\omega_1+\delta)
}{
(s-\delta)(\delta+2i\gamma)
},
\qquad
I_2
=
\int\frac{d\delta}{2\pi}
\frac{
\tilde{\Pi}_0^R(\omega_4)-\tilde{\Pi}_0^R(\omega_4-\delta)
}{
(s-\delta+2i\gamma)\delta
},
\label{eqn:ABB_I}
\end{equation}
and
\begin{equation}
I_{\mathrm{sp}}
=
\left[
\tilde{\Pi}_0^A(\omega_1)-\tilde{\Pi}_0^R(\omega_4)
\right]
\int\frac{d\delta}{2\pi}
\frac{1}{
(s-\delta+2i\gamma)(\delta+2i\gamma)
}.
\label{eqn:ABB_Isp}
\end{equation}

We first consider the $I_1$ term in Eq.~\eqref{eqn:ABB_I}. The apparent pole at $\delta=s$ is removable since $\omega_1+s=\omega_4$, and the only remaining explicit pole is therefore $\delta=-2i\gamma$, which lies in the lower half-plane. Since $\tilde{\Pi}_0^R(\omega_1+\delta)$ is analytic for $\operatorname{Im}\delta>0$, the contour may be closed in the upper half-plane, where no singularities are enclosed, and hence $I_1=0$.

The same reasoning applies to $I_2$. The apparent pole at $\delta=0$ is removable, and the remaining explicit pole is $\delta=s+2i\gamma$, which lies in the upper half-plane. Moreover, $\tilde{\Pi}_0^R(\omega_4-\delta)$ is analytic for $\operatorname{Im}\delta<0$. We may therefore close the contour in the lower half-plane and obtain $I_2=0$.

The rephasing sector thus isolates the spectral part of the ordered square. The remaining integral contains two genuine poles, $\delta_-=-2i\gamma$, and $\delta_+=s+2i\gamma$, which lie on opposite sides of the real axis. Closing the contour in the lower half-plane gives
\begin{equation}
K_{\rm R}^R(\omega_1,\omega_4)
=
-i\,
\frac{
\tilde{\Pi}_0^A(\omega_1)-\tilde{\Pi}_0^R(\omega_4)
}{
\omega_4-\omega_1+4i\gamma
}.
\label{eqn:ABB_projected_kernel}
\end{equation}

We finally Fourier transform with respect to the two physical delays. With our convention, the two-dimensional spectrum is
\begin{equation}
\mathcal S_{0,\rm R}^R(\omega_\tau,\omega_t)
=
\int_0^\infty d\tau
\int_0^\infty dt\,
e^{i\omega_\tau\tau}
e^{i\omega_t t}
\mathcal S_{0,\rm R}^R(t,\tau).
\label{eqn:ABB_2D_transform}
\end{equation}
Substituting Eq.~\eqref{eqn:ABB_kernel} gives
\begin{equation}
\begin{aligned}
\mathcal S_{0,\rm R}^R(\omega_\tau,\omega_t)
=
\int\frac{d\omega_1}{2\pi}
\frac{d\omega_4}{2\pi}\,
K_{\rm R}^R(\omega_1,\omega_4)
\frac{i}{\omega_\tau+\omega_1+i0^+}
\frac{i}{\omega_t-\omega_4+i0^+}.
\end{aligned}
\label{eqn:ABB_causal_transform}
\end{equation}
The one-sided time integrations therefore generate causal Cauchy kernels rather than delta functions since $\tau,t\geq0$. As a function of $\omega_1$, the projected kernel in Eq.~\eqref{eqn:ABB_projected_kernel} is analytic in the lower half-plane, where $\tilde{\Pi}_0^A(\omega_1)$ is analytic, while the explicit denominator has its pole at $\omega_1=\omega_4+4i\gamma$, in the upper half-plane. The first Cauchy kernel has a pole at $\omega_1=-\omega_\tau-i0^+$. Closing the $\omega_1$ contour in the lower half-plane therefore gives
\begin{equation}
\int\frac{d\omega_1}{2\pi}\,
K_{\rm R}^R(\omega_1,\omega_4)
\frac{i}{\omega_\tau+\omega_1+i0^+}
=
K_{\rm R}^R(-\omega_\tau,\omega_4).
\label{eqn:ABB_omega1_projection}
\end{equation}
The remaining kernel is analytic in the upper $\omega_4$ half-plane, where $\tilde{\Pi}_0^R(\omega_4)$ is analytic, while the explicit pole of Eq.~\eqref{eqn:ABB_projected_kernel} lies at $\omega_4=-\omega_\tau-4i\gamma$, in the lower half-plane. The second Cauchy kernel has a pole at $\omega_4=\omega_t+i0^+$. Closing the $\omega_4$ contour in the upper half-plane therefore yields $\mathcal S_{0,\rm R}^R(\omega_\tau,\omega_t)=K_{\rm R}^R(-\omega_\tau,\omega_t)$. Thus the two experimental frequency axes are related to the arrow-oriented frequencies of the square according to $\omega_1=-\omega_\tau$, and $\omega_4=\omega_t$. The first-delay frequency $\omega_\tau$ is therefore the sign-reversed frequency associated with the first pulse, whereas $\omega_t$ is the emitted frequency during the detection interval. Using Eq.~\eqref{eqn:ABB_projected_kernel}, the two-dimensional ABB spectrum is finally
\begin{equation}
\mathcal S_{0,\rm R}^R(\omega_\tau,\omega_t)
=
i\,
\frac{
\tilde{\Pi}_0^R(\omega_t)-\tilde{\Pi}_0^A(-\omega_\tau)
}{
\omega_t+\omega_\tau+4i\gamma
}. 
\end{equation}
which corresponds to Eq.~\eqref{eqn:ABB_echo_map} of the main text.

\subsection{Nonrephasing response}
\label{App: NR_calculation}

We now consider the nonrephasing contribution (NR) obtained by interchanging the external frequencies $\omega_1$ and $\omega_2$ in Eq.~\eqref{eqn:bubble_square_relation}. The corresponding frequency-conservation condition is $\omega_1+\omega_3=\omega_2+\omega_4$. The ABB chronology remains
\begin{equation}
t_1=0,
\qquad
t_2=t_3=\tau,
\qquad
t_4=\tau+t,
\qquad
\tau>0,
\qquad
t>0.
\label{eqn:NR_ABB_times}
\end{equation}
After interchanging $1$ and $2$, legs $1$ and $3$ are incoming while legs $2$ and $4$ are outgoing. The external Fourier phase is therefore
\begin{equation}
e^{+i\omega_1t_1}
e^{-i\omega_2t_2}
e^{+i\omega_3t_3}
e^{-i\omega_4t_4}
=
e^{-i\omega_1\tau}
e^{-i\omega_4t},
\label{eqn:NR_ABB_phase}
\end{equation}
where we have used $\omega_1+\omega_3=\omega_2+\omega_4$. Thus, in contrast with the rephasing contribution, the first-delay coherence and the emitted coherence enter with the same Fourier orientation. The corresponding ABB signal can be written as
\begin{equation}
\mathcal S_{0,\rm NR}^R(t,\tau)
=
\int\frac{d\omega_1}{2\pi}
\frac{d\omega_2}{2\pi}
\frac{d\omega_3}{2\pi}\,
e^{-i\omega_1\tau}
e^{-i\omega_4t}
\mathcal S^R_0
(\omega_2,\omega_1,\omega_3).
\label{eqn:NR_ABB_time_response}
\end{equation}
As in the rephasing sector, the equality $t_2=t_3$ removes one independent frequency from the Fourier phase. Introducing $\delta\equiv\omega_2-\omega_1$, frequency conservation gives $\omega_2=\omega_1+\delta$, and $\omega_3=\omega_4+\delta$. The ABB signal therefore becomes
\begin{equation}
\mathcal S_{0,\mathrm{NR}}^R(t,\tau)
=
\int\frac{d\omega_1}{2\pi}
\frac{d\omega_4}{2\pi}\,
e^{-i\omega_1\tau}
e^{-i\omega_4t}
K_{\mathrm{NR}}^R(\omega_1,\omega_4),
\quad
\text{where}
\quad
K_{\mathrm{NR}}^R(\omega_1,\omega_4)
\equiv
\int\frac{d\delta}{2\pi}\,
\mathcal S_{0,\mathrm{NR}}^R
\left(
\omega_1,
\omega_1+\delta,
\omega_4+\delta
\right),
\label{eqn:NR_ABB_time_kernel}
\end{equation}
and $S_{0,\mathrm{NR}}^R(\omega_1,\omega_2,\omega_3) \equiv S_0^R(\omega_2,\omega_1,\omega_3) $. The frequency routing is now qualitatively different from the rephasing case. Defining $s\equiv\omega_4-\omega_1$, one finds $\omega_1-\omega_2=-\delta$, whereas $\omega_3-\omega_2=s$. Thus only one of the two detunings appearing in Eq.~\eqref{eqn:bubble_square_relation} depends on the projected frequency $\delta$, resulting in
\begin{equation}
K_{\rm NR}^R(\omega_1,\omega_4)
=
J_1+J_2+J_3,
\label{eqn:NR_ABB_decomposition}
\end{equation}
where
\begin{equation}
J_1
=
\frac{
\tilde{\Pi}_0^R(\omega_4)-\tilde{\Pi}_0^R(\omega_1)
}{s}
\int\frac{d\delta}{2\pi}
\frac{1}{-\delta+2i\gamma},
\qquad
J_2
=
\frac{1}{s+2i\gamma}
\int\frac{d\delta}{2\pi}
\frac{
\tilde{\Pi}_0^R(\omega_4)-\tilde{\Pi}_0^R(\omega_4+\delta)
}{-\delta},
\label{eqn:NR_ABB_J12}
\end{equation}
and
\begin{equation}
J_3
=
\frac{1}{s+2i\gamma}
\int\frac{d\delta}{2\pi}
\frac{
\tilde{\Pi}_0^A(\omega_1+\delta)-\tilde{\Pi}_0^R(\omega_4)
}{
-\delta+2i\gamma
}.
\label{eqn:NR_ABB_J3}
\end{equation}
In contrast with the rephasing projection, the individual integrands in Eqs.~\eqref{eqn:NR_ABB_J12} and \eqref{eqn:NR_ABB_J3} fall only as $1/\delta$ at large $|\delta|$. This reflects the fact that the two interactions with the $B$ pulse occur at the same time. In the impulsive limit the integrals are therefore understood with the symmetric equal-time prescription, equivalent to $\theta(0)=1/2$. A finite pulse envelope provides the corresponding physical regularization.

The first integral is immediately
\begin{equation}
\int_{-\infty}^{\infty}\frac{d\delta}{2\pi}
\frac{1}{-\delta+2i\gamma}
=
-\frac{i}{2},
\quad 
\text{and hence}
\quad
J_1
=
-\frac{i}{2}
\frac{
\tilde{\Pi}_0^R(\omega_4)-\tilde{\Pi}_0^R(\omega_1)
}{
\omega_4-\omega_1
}.
\label{eqn:NR_ABB_J1_result}
\end{equation}
For $J_2$, the apparent pole at $\delta=0$ is removable and $\tilde{\Pi}_0^R(\omega_4+\delta)$ is analytic in the upper $\delta$ half-plane. There are therefore no finite poles in the upper half-plane. The integrand approaches $\tilde{\Pi}_0^R(\omega_4)/(-\delta)$ for large $|\delta|$, so the large semicircle gives the equal-time contribution. Closing the contour in the upper half-plane yields
\begin{equation}
J_2
=
\frac{i}{2}
\frac{
\tilde{\Pi}_0^R(\omega_4)
}{
s+2i\gamma
}.
\label{eqn:NR_ABB_J2_result}
\end{equation}

For $J_3$, $\tilde{\Pi}_0^A(\omega_1+\delta)$ is analytic in the lower $\delta$ half-plane, while the explicit pole at $\delta=2i\gamma$ lies in the upper half-plane. We may therefore close the contour below. The large-$|\delta|$ behavior again gives the equal-time contribution and one obtains
\begin{equation}
J_3
=
\frac{i}{2}
\frac{
\tilde{\Pi}_0^R(\omega_4)
}{
s+2i\gamma
}.
\label{eqn:NR_ABB_J3_result}
\end{equation}
Combining the three terms, the projected nonrephasing kernel is
\begin{equation}
K_{\rm NR}^R(\omega_1,\omega_4)
=
-\frac{i}{2}
\frac{
\tilde{\Pi}_0^R(\omega_4)-\tilde{\Pi}_0^R(\omega_1)
}{
\omega_4-\omega_1
}
+
i
\frac{
\tilde{\Pi}_0^R(\omega_4)
}{
\omega_4-\omega_1+2i\gamma
}.
\label{eqn:NR_ABB_projected_kernel}
\end{equation}

We finally Fourier transform with respect to the two physical delays,
\begin{equation}
\mathcal S_{0,\rm NR}^R(\omega_\tau,\omega_t)
=
\int_0^\infty d\tau
\int_0^\infty dt\,
e^{i\omega_\tau\tau}
e^{i\omega_t t}
\mathcal S_{0,\rm NR}^R(t,\tau).
\label{eqn:NR_ABB_2D_transform}
\end{equation}
Substituting Eq.~\eqref{eqn:NR_ABB_time_kernel} gives
\begin{equation}
\mathcal S_{0,\rm NR}^R(\omega_\tau,\omega_t)
=
\int\frac{d\omega_1}{2\pi}
\frac{d\omega_4}{2\pi}\,
K_{\rm NR}^R(\omega_1,\omega_4)
\frac{i}{\omega_\tau-\omega_1+i0^+}
\frac{i}{\omega_t-\omega_4+i0^+}.
\label{eqn:NR_ABB_causal_transform}
\end{equation}
Since $\tau,t\geq0$, the one-sided time integrations again generate causal Cauchy kernels rather than delta functions.

We first perform the $\omega_1$ integration. The first Cauchy kernel has its pole at $\omega_1=\omega_\tau+i0^+$. In addition, the second term of Eq.~\eqref{eqn:NR_ABB_projected_kernel} has a pole at $\omega_1=\omega_4+2i\gamma$, also in the upper half-plane. Closing the contour above therefore picks both poles. The contribution from the Cauchy pole is $K_{\rm NR}^R(\omega_\tau,\omega_4)$, while the second pole produces a term which exactly cancels the second term in Eq.~\eqref{eqn:NR_ABB_projected_kernel}. We are therefore left with
\begin{equation}
\int\frac{d\omega_1}{2\pi}\,
K_{\rm NR}^R(\omega_1,\omega_4)
\frac{i}{\omega_\tau-\omega_1+i0^+}
=
-\frac{i}{2}
\frac{
\tilde{\Pi}_0^R(\omega_4)-\tilde{\Pi}_0^R(\omega_\tau)
}{
\omega_4-\omega_\tau
}.
\label{eqn:NR_ABB_omega1_integral}
\end{equation}

The remaining divided difference is analytic in the upper $\omega_4$ half-plane. The apparent singularity at $\omega_4=\omega_\tau$ is removable, since
\begin{equation}
\lim_{\omega_4\rightarrow\omega_\tau}
\frac{
\tilde{\Pi}_0^R(\omega_4)-\tilde{\Pi}_0^R(\omega_\tau)
}{
\omega_4-\omega_\tau
}
=
\partial_\omega\tilde{\Pi}_0^R(\omega_\tau).
\label{eqn:NR_ABB_divided_difference}
\end{equation}
The second Cauchy kernel has its pole at $\omega_4=\omega_t+i0^+$. Closing the $\omega_4$ contour in the upper half-plane therefore gives
\begin{equation}
\mathcal S_{0,\rm NR}^R(\omega_\tau,\omega_t)
=
-\frac{i}{2}
\frac{
\tilde{\Pi}_0^R(\omega_t)-\tilde{\Pi}_0^R(\omega_\tau)
}{
\omega_t-\omega_\tau
},
\end{equation}
corresponding to Eq.~\eqref{eqn:NR_ABB_map} of the main text.

\section{Generalities and limitations}
\label{sec:generalities}

% In this section, we discuss the generality of the results obtained in the previous sections. In particular, we examine the effects of momentum-dependent vertices, momentum dependent homogeneous broadening, and a frequency-dependent self-energy. We identify which of our previous conclusions remain valid in these more general situations and discuss the limitations of the approximations introduced above.

In this section, we examine the generality of the preceding results by allowing for momentum-dependent homogeneous broadening and a frequency-dependent self-energy. We first show that momentum-dependent linewidths preserve the reduction at each momentum, Eq.~\eqref{eqn: general_gamma_k_relation}, and yield a superposition of rephasing contributions with distinct homogeneous widths. We then show how a genuinely frequency-dependent self-energy obstructs the reduction to ordinary pair bubbles and identify the quasiparticle regime in which an approximate construction remains possible. 
% Finally, we illustrate the formalism through interband transitions, establishing the equivalence between number-conserving electron scattering and electron–hole pair creation. This example clarifies why total particle-number conservation and diagrammatic arrow orientation alone do not determine whether rephasing occurs.

\subsection{Momentum-dependent quasiparticle linewidth}
\label{sec: mom_dep_gamma}

We have so far assumed a momentum-independent quasiparticle linewidth 
$\gamma$. We now consider the more general case
\begin{equation}
G_{\mathbf{k}}^{R/A}(\omega)
=
\frac{1}
{\omega-\xi_{\mathbf{k}}\pm i\gamma_{\mathbf{k}}},
\end{equation}
while retaining a frequency-independent self-energy, and assume inversion symmetry of the damping rate, $\gamma_{-\mathbf{k}}=\gamma_{\mathbf{k}}$. The corresponding pair bubble becomes
\begin{equation}
\Pi_0^{R/A}(\omega)
=
2
\int_{\mathbf{k}}
\frac{\abs{f_\mathbf{k}}^2\,
F_{\eta,\mathbf{k}}
}{
\omega-2\xi_{\mathbf{k}}\pm 2i\gamma_{\mathbf{k}}
} = \int_{\mathbf{k}}\abs{f_\mathbf{k}}^2\,\pi_{\mathbf{k}}^{R/A}(\omega).
\end{equation}
Importantly, momentum dependence of the linewidth does not affect the 
frequency-space identities used to reduce the square at fixed momentum. 
For example,
\begin{equation}
G_{\mathbf{k}}^{R}(x)
G_{\mathbf{k}}^{R}(y)
=
\frac{
G_{\mathbf{k}}^{R}(y)
-
G_{\mathbf{k}}^{R}(x)
}{
x-y
},
\qquad \text{or} \qquad
G_{\mathbf{k}}^{R}(x)
G_{\mathbf{k}}^{A}(y)
=
\frac{
G_{\mathbf{k}}^{A}(y)
-
G_{\mathbf{k}}^{R}(x)
}{
x-y+2i\gamma_{\mathbf{k}}
}.
\end{equation}
Thus, the complete bubble--square reduction can still be carried out 
independently for every $\mathbf{k}$. Introducing the momentum-resolved pair bubble the analogue of Eq.~\eqref{eqn:bubble_square_relation} takes the form
\begin{equation}
\mathcal{S}_0^{R}(\omega_1,\omega_2,\omega_3)=
\int_{\mathbf{k}}\abs{f_\mathbf{k}}^4
\frac{
\pi_{\mathbf{k}}^{R}(\omega_4)
-
\pi_{\mathbf{k}}^{R}(\omega_2)
}{
\Delta_{31}
\left(
\Delta_{21}+2i\gamma_{\mathbf{k}}
\right)
}+
\int_{\mathbf{k}}\abs{f_\mathbf{k}}^4
\frac{
\pi_{\mathbf{k}}^{R}(\omega_4)
-
\pi_{\mathbf{k}}^{R}(\omega_3)
}{
\left(
\Delta_{31}+2i\gamma_{\mathbf{k}}
\right)
\Delta_{21}
}+
\int_{\mathbf{k}}\abs{f_\mathbf{k}}^4
\frac{
\pi_{\mathbf{k}}^{A}(\omega_1)
-
\pi_{\mathbf{k}}^{R}(\omega_4)
}{
\left(
\Delta_{31}+2i\gamma_{\mathbf{k}}
\right)
\left(
\Delta_{21}+2i\gamma_{\mathbf{k}}
\right)
}.
\label{eqn: general_gamma_k_relation}
\end{equation}

Momentum-dependent damping therefore does not obstruct the decomposition 
of the four-propagator square into two-propagator objects. It does, however, 
prevent the result from being written solely in terms of the momentum-integrated 
pair bubble $\Pi^{R/A}(\omega)$. In the momentum-independent case, the factors 
$\left(\Delta+2i\gamma\right)^{-1}$ can be taken outside the momentum integral, 
leading directly to Eq.~\eqref{eqn:bubble_square_relation}. For $\gamma_{\mathbf{k}}$, these factors remain 
inside the integral and act as additional momentum-dependent weights.

Note that the on-shell approximation used for the equilibrium Keldysh 
component now has to hold at each relevant momentum, $\gamma_{\mathbf{k}}
\left|
\partial_{\omega}F_{\eta}(\omega)
\right|_{\omega=\xi_{\mathbf{k}}}
\ll
\left|
F_{\eta,\mathbf{k}}
\right|$. Within this regime, momentum dependence of the linewidth is therefore a 
relatively mild generalization of the present construction. In contrast, 
a genuine frequency dependence of the self-energy modifies the 
frequency-space identities themselves and constitutes a more fundamental 
obstruction to the bubble--square reduction, as we will see below.

% \subsubsection{Two-dimensional rephasing response}

It is useful to examine how the momentum-dependent quasiparticle linewidth
appears in the two-dimensional rephasing response. Since the reduction above
holds independently at every momentum, the frequency projection introduced
in Eq.~\eqref{eqn:ABB_kernel} can likewise be performed before the
momentum integral. We therefore write
\begin{equation}
\mathcal{K}_{R}^{R}(\omega_1,\omega_4)
=
\int_{\mathbf{k}}
\int\frac{d\delta}{2\pi}
\,
\mathcal{S}_{0,\mathbf{k}}^{R}
\left(
\omega_1,
\omega_1+\delta,
\omega_4-\delta
\right),
\end{equation}
where $\omega_2=\omega_1+\delta$, and $\omega_3=\omega_4-\delta$. Defining, as before, $s=\omega_4-\omega_1$, the first two contributions to the square become
\begin{equation}
I_{1,\mathbf{k}}
=
\int\frac{d\delta}{2\pi}\abs{f_\mathbf{k}}^4
\frac{
\pi_{\mathbf{k}}^{R}(\omega_4)
-
\pi_{\mathbf{k}}^{R}(\omega_1+\delta)
}{
(s-\delta)
(\delta+2i\gamma_{\mathbf{k}})
},
\qquad
\text{and}
\qquad
I_{2,\mathbf{k}}
=
\int\frac{d\delta}{2\pi}\abs{f_\mathbf{k}}^4
\frac{
\pi_{\mathbf{k}}^{R}(\omega_4)
-
\pi_{\mathbf{k}}^{R}(\omega_4-\delta)
}{
(s-\delta+2i\gamma_{\mathbf{k}})
\delta
}.
\end{equation}
The same analyticity arguments employed for the momentum-independent
linewidth apply separately for every $\mathbf{k}$. The apparent poles at
$\delta=s$ and $\delta=0$ are removable, while the remaining explicit poles
lie in the half-plane opposite to that in which the corresponding retarded
pair bubble is analytic. Consequently, $I_{1,\mathbf{k}}=I_{2,\mathbf{k}}=0$.

The rephasing projection therefore again isolates the spectral contribution.
At fixed momentum this contribution is
\begin{equation}
I_{{\rm sp},\mathbf{k}}
=
\abs{f_\mathbf{k}}^4\left[
\pi_{\mathbf{k}}^{A}(\omega_1)
-
\pi_{\mathbf{k}}^{R}(\omega_4)
\right]
\int\frac{d\delta}{2\pi}
\frac{1}{
(s-\delta+2i\gamma_{\mathbf{k}})
(\delta+2i\gamma_{\mathbf{k}})
}.
\end{equation}
The two poles, $\delta_{-}=-2i\gamma_{\mathbf{k}}$ and $\delta_{+}=s+2i\gamma_{\mathbf{k}}$, lie on opposite sides of the real axis. Performing the contour integral either closing the contour from above or below gives
\begin{equation}
\mathcal{K}_{R}^{R}(\omega_1,\omega_4)
=
-i
\int_{\mathbf{k}}\abs{f_\mathbf{k}}^4
\frac{
\pi_{\mathbf{k}}^{A}(\omega_1)
-
\pi_{\mathbf{k}}^{R}(\omega_4)
}{
\omega_4-\omega_1+4i\gamma_{\mathbf{k}}
}.
\label{eqn:gamma_k_projected_kernel}
\end{equation}

We now perform the one-sided Fourier transforms with respect to the two
physical delays. Since the analyticity properties used in the
momentum-independent case hold separately for every $\mathbf{k}$, the two
Cauchy integrations again evaluate the projected kernel at $\omega_1=-\omega_\tau$ and $\omega_4=\omega_t$. The complete two-dimensional rephasing response is therefore
\begin{equation}
\mathcal{S}_{R}^{R}(\omega_\tau,\omega_t)
=
i
\int_{\mathbf{k}} \abs{f_\mathbf{k}}^4
\frac{
\pi_{\mathbf{k}}^{R}(\omega_t)
-
\pi_{\mathbf{k}}^{A}(-\omega_\tau)
}{
\omega_t+\omega_\tau+4i\gamma_{\mathbf{k}}
}.
\label{eqn:gamma_k_2d_map}
\end{equation}

Using the explicit momentum-resolved pair bubble, Eq.~\eqref{eqn:gamma_k_2d_map} can be simplified further using
\begin{equation}
\pi_{\mathbf{k}}^{R}(\omega_t)
-
\pi_{\mathbf{k}}^{A}(-\omega_\tau)
=
-
2F_{\eta,\mathbf{k}}\,
\frac{
\omega_t+\omega_\tau+4i\gamma_{\mathbf{k}}
}{
\left(
\omega_t-2\xi_{\mathbf{k}}+2i\gamma_{\mathbf{k}}
\right)
\left(
-\omega_\tau-2\xi_{\mathbf{k}}-2i\gamma_{\mathbf{k}}
\right)
}.
\end{equation}
The factor appearing in the numerator cancels exactly against the
rephasing denominator in Eq.~\eqref{eqn:gamma_k_2d_map}. We therefore obtain
\begin{equation}
\mathcal{S}_{R}^{R}(\omega_\tau,\omega_t)
=
-2i
\int_{\mathbf{k}}
\frac{\abs{f_\mathbf{k}}^4\,
F_{\eta,\mathbf{k}}
}{
\left(
\omega_t-2\xi_{\mathbf{k}}+2i\gamma_{\mathbf{k}}
\right)
\left(
-\omega_\tau-2\xi_{\mathbf{k}}-2i\gamma_{\mathbf{k}}
\right)
}.
\label{eqn:gamma_k_2d_map_final}
\end{equation}
Thus the full rephasing spectrum is a superposition of momentum-resolved
echo contributions, each centered at the pair excitation energy
$2\xi_{\mathbf{k}}$ and characterized by its own homogeneous linewidth
$2\gamma_{\mathbf{k}}$.

\subsection{Frequency-dependent self-energy}
\label{sec: freq_sigma}

We have so far assumed a frequency-independent quasiparticle self-energy in Eq.~\eqref{eqn: free_propagator}. We now consider the more general propagator containing a general frequency-dependent self-energy
\begin{equation}
G_{\mathbf{k}}^{R/A}(\omega)
=
\frac{1}
{\omega-\xi_{\mathbf{k}}
-\Sigma_{\mathbf{k}}^{R/A}(\omega)}.
\end{equation}
A genuine frequency dependence of the self-energy constitutes a more fundamental modification of the bubble--square relation than either a momentum-dependent vertex or a momentum-dependent linewidth, discussed previously. The reason is that it modifies the frequency-space identities used to reduce products of propagators we have extensively used until this point. For two propagators one finds exactly
\begin{equation}
G_{\mathbf{k}}^{R}(x)
G_{\mathbf{k}}^{R}(y)
=
\frac{
G_{\mathbf{k}}^{R}(y)
-
G_{\mathbf{k}}^{R}(x)
}{
(x-y)
-
\left[
\Sigma_{\mathbf{k}}^{R}(x)
-
\Sigma_{\mathbf{k}}^{R}(y)
\right]
} 
\qquad \text{and} \qquad
G_{\mathbf{k}}^{R}(x)
G_{\mathbf{k}}^{A}(y)
=
\frac{
G_{\mathbf{k}}^{A}(y)
-
G_{\mathbf{k}}^{R}(x)
}{
(x-y)
-
\Sigma_{\mathbf{k}}^{R}(x)
+
\Sigma_{\mathbf{k}}^{A}(y)
}.
\end{equation}
The crucial difference for a dynamical self-energy is that the additional terms in the denominators depend on the internal loop frequency. For example, using the frequency routing introduced above, $x-y=\omega_3-\omega_1$, whereas
\begin{equation}
\Sigma_{\mathbf{k}}^{R}(x)
-
\Sigma_{\mathbf{k}}^{R}(y)
=
\Sigma_{\mathbf{k}}^{R}
\left(
\nu+\omega_3-\omega_1
\right)
-
\Sigma_{\mathbf{k}}^{R}(\nu).
\end{equation}
Although $x-y$ is independent of the internal frequency $\nu$, the self-energy difference is, in general, not. Consequently, the denominator can no longer be pulled outside the frequency integral. The same issue occurs in the mixed retarded--advanced terms through combinations of the form $\Sigma_{\mathbf{k}}^{R}(x)-\Sigma_{\mathbf{k}}^{A}(y)$.

The original bubble--square relation can be nevertheless recovered approximately in the quasiparticle regime. If the self-energy can be linearized over the frequency window sampled by the response, the single-particle propagator takes the pole form
\begin{equation}
G_{\mathbf{k}}^{R/A}(\omega)
\simeq
\frac{
Z_{\mathbf{k}}
}{
\omega-E_{\mathbf{k}}
\pm i\Gamma_{\mathbf{k}}
},
\end{equation}
where $E_{\mathbf{k}}$ is the renormalized quasiparticle dispersion, $Z_{\mathbf{k}}$ the quasiparticle residue, and $\Gamma_{\mathbf{k}}$ the quasiparticle linewidth. Products of propagators then again satisfy simple finite-difference identities, for example
\begin{equation}
G_{\mathbf{k}}^{R}(x)
G_{\mathbf{k}}^{R}(y)
\simeq
Z_{\mathbf{k}}
\frac{
G_{\mathbf{k}}^{R}(y)
-
G_{\mathbf{k}}^{R}(x)
}{
x-y
},
\qquad \text{and} \qquad
G_{\mathbf{k}}^{R}(x)
G_{\mathbf{k}}^{A}(y)
\simeq
Z_{\mathbf{k}}
\frac{
G_{\mathbf{k}}^{A}(y)
-
G_{\mathbf{k}}^{R}(x)
}{
x-y+2i\Gamma_{\mathbf{k}}
}.
\end{equation}
Thus, within a well-defined quasiparticle regime, the effect of a dynamical
self-energy can be absorbed approximately into renormalized dispersions,
residues, and linewidths. The bubble--square construction then survives in
the generalized form discussed above for momentum-dependent weights and
linewidths.

A genuinely dynamical self-energy, for which the variation of
$\Sigma_{\mathbf{k}}^{R/A}(\omega)$ across the relevant frequency window
cannot be neglected, lies outside this quasiparticle reduction. In this
regime, the nonlinear square contains information about the frequency
dependence of the single-particle self-energy that is not encoded in the
ordinary linear pair susceptibility alone.

\section{Applications}
\label{App: Applications}

\subsection{Application to interband scattering}
\label{app:interband_pair}

Zero-momentum interband transitions provide a nontrivial application of our formalism and clarify the algebraic origin of the rephasing physics. Although the probe conserves the total electron number, it connects bands whose occupations generally differ. Their propagators therefore need not admit a common, frequency-independent diagonalization in Keldysh space, allowing mixed retarded--advanced contributions to survive. We first describe these transitions as electron scattering between bands and then establish the equivalence of this description to electron--hole pair creation.

\subsubsection{Number-conserving interband representation}

We consider noninteracting conduction and valence bands,
\begin{equation}
H_0
=
\sum_{\mathbf k}
\left[
\epsilon_{c,\mathbf k}c_{\mathbf k}^\dagger c_{\mathbf k}
+
\epsilon_{v,\mathbf k}v_{\mathbf k}^\dagger v_{\mathbf k}
\right],
\end{equation}
where the energies are measured relative to the chemical potential. The external probe couples through
\begin{equation}
H_{\mathrm{probe}}(t)
=
-J(t)P^\dagger-J^*(t)P,
\qquad
P^\dagger
=
\sum_{\mathbf k}d_{\mathbf k}c_{\mathbf k}^\dagger v_{\mathbf k}.
\end{equation}
The operator $P^\dagger$ transfers an electron from the valence band to the conduction band, with matrix element $d_{\mathbf k}$. The corresponding square has consistently oriented electron arrows, while the band labels alternate around the loop. For each band $a=c,v$, we introduce $G_{a,\mathbf k}^{R/A}(\nu) = \left[\nu-\epsilon_{a,\mathbf k}\pm i\gamma_{a,\mathbf k}\right]^{-1}$, with frequency-independent linewidths. Within the on-shell approximation, the causal matrices are
\begin{equation}
\mathbb G_{a,\mathbf k}(\nu)
=
\begin{pmatrix}
G_{a,\mathbf k}^R(\nu)
&
F_{a,\mathbf k}\Delta_{a,\mathbf k}(\nu)
\\
0&G_{a,\mathbf k}^A(\nu)
\end{pmatrix},
\qquad
F_{a,\mathbf k}=1-2n_{a,\mathbf k},
\qquad
\Delta_{a,\mathbf k}=G_{a,\mathbf k}^R-G_{a,\mathbf k}^A.
\end{equation}
Each matrix admits a frequency-independent diagonalization, see Eq.~\eqref{eqn:similarity_G},
\begin{equation}
\mathbb G_{a,\mathbf k}(\nu)
=
U_{a,\mathbf k}
\begin{pmatrix}
G_{a,\mathbf k}^R(\nu)&0\\
0&G_{a,\mathbf k}^A(\nu)
\end{pmatrix}
U_{a,\mathbf k}^{-1},
\qquad
U_{a,\mathbf k}
=
\begin{pmatrix}
1&-F_{a,\mathbf k}\\
0&1
\end{pmatrix}.
\end{equation}
However, commuting the matrix propagators of different bands introduces the product
\begin{equation}
U_{c,\mathbf k}^{-1}U_{v,\mathbf k}
=
\begin{pmatrix}
1&F_{c,\mathbf k}-F_{v,\mathbf k}\\
0&1
\end{pmatrix},
\end{equation}
which mixes the retarded and advanced sectors when the occupations differ and direct multiplication results in
\begin{equation}
\left[
\mathbb G_{c,\mathbf k}(\nu),
\mathbb G_{v,\mathbf k}(\nu')
\right]
=
\left(F_{v,\mathbf k}-F_{c,\mathbf k}\right)
\Delta_{c,\mathbf k}(\nu)\Delta_{v,\mathbf k}(\nu')
\begin{pmatrix}
0&1\\
0&0
\end{pmatrix}.
\label{eq:interband_commutator}
\end{equation}
Reordering the propagators generates spectral terms analogous to those in Eq.~\eqref{eqn: cross_commutator}, which support rephasing. If the two bands have equal occupations at each momentum, however, the commutator vanishes and a common causal diagonalization is possible. Note, however, that in this equal-occupation limit, both the linear and third-order responses vanish within the present approximation. Since common causal diagonalization leaves only purely retarded or purely advanced products, and all electron propagators have arguments of the form $\nu+\alpha_j$, their internal-frequency integrals vanish. This argument does not apply to the regular particle--particle contribution, discussed in the main text, whose opposite frequency routings place poles in both half-planes.

\subsubsection{Electron--hole representation}

We now perform a particle--hole transformation on the valence band, $h_{-\mathbf k}^\dagger=v_{\mathbf k}$, and $h_{-\mathbf k}=v_{\mathbf k}^\dagger$. The Hamiltonian and probe operator become
\begin{equation}
H_0
=
\mathrm{const.}
+
\sum_{\mathbf k}
\left[
\epsilon_{c,\mathbf k}c_{\mathbf k}^\dagger c_{\mathbf k}
+
\epsilon_{h,-\mathbf k}h_{-\mathbf k}^\dagger h_{-\mathbf k}
\right],
\qquad
\epsilon_{h,-\mathbf k}=-\epsilon_{v,\mathbf k},
\qquad
\text{and}
\qquad
P^\dagger
=
\sum_{\mathbf k}d_{\mathbf k}c_{\mathbf k}^\dagger h_{-\mathbf k}^\dagger.
\end{equation}
The same transition is now represented as electron--hole pair creation. As in the pair channel discussed earlier, the hole line enters through the reversed causal matrix,
\begin{equation}
\overline{\mathbb G}_{h,-\mathbf k}(\omega)
\equiv
\sigma_x\mathbb G_{h,-\mathbf k}^{T}(\omega)\sigma_x
=
\begin{pmatrix}
G_{h,-\mathbf k}^A(\omega)
&
F_{h,-\mathbf k}\Delta_{h,-\mathbf k}(\omega)
\\
0&G_{h,-\mathbf k}^R(\omega)
\end{pmatrix}.
\end{equation}
Its commutator with the conduction-electron matrix is
\begin{equation}
\left[
\mathbb G_{c,\mathbf k}(\nu),
\overline{\mathbb G}_{h,-\mathbf k}(\omega)
\right]
=
\left(F_{c,\mathbf k}+F_{h,-\mathbf k}\right)
\Delta_{c,\mathbf k}(\nu)\Delta_{h,-\mathbf k}(\omega)
\begin{pmatrix}
0&1\\
0&0
\end{pmatrix}.
\label{eq:electron_hole_commutator}
\end{equation}
The coefficient is now the sum of the electron and hole distribution factors. Equal electron and hole occupations therefore do not generally imply commutativity: in particular, an empty conduction band and a filled valence band correspond to the electron--hole vacuum, for which $F_{c,\mathbf k}=F_{h,-\mathbf k}=1$.

The equivalence of the two causal descriptions follows from
\begin{equation}
G_{h,-\mathbf k}^{R}(-\nu)
=
-G_{v,\mathbf k}^{A}(\nu),
\qquad
G_{h,-\mathbf k}^{A}(-\nu)
=
-G_{v,\mathbf k}^{R}(\nu).
\end{equation}
Together with $F_{h,-\mathbf k}=-F_{v,\mathbf k}$, these relations imply the matrix identity
\begin{equation}
\mathbb G_{v,\mathbf k}(\nu)
=
-\overline{\mathbb G}_{h,-\mathbf k}(-\nu).
\label{eq:electron_hole_matrix_identity}
\end{equation}
Consequently,
\begin{equation}
\left[
\mathbb G_{c,\mathbf k}(\nu),
\mathbb G_{v,\mathbf k}(\nu')
\right]
=
-
\left[
\mathbb G_{c,\mathbf k}(\nu),
\overline{\mathbb G}_{h,-\mathbf k}(-\nu')
\right],
\end{equation}
establishing the equivalence of Eqs.~\eqref{eq:interband_commutator} and~\eqref{eq:electron_hole_commutator}. The same identity also relates the square diagrams directly. Denoting the four internal line frequencies by $\nu_j$ and suppressing momentum labels, we have
\begin{equation}
\operatorname{Tr}_{\mathrm K}
\left[
\sigma_x
\mathbb G_c(\nu_1)
\mathbb G_v(\nu_2)
\mathbb G_c(\nu_3)
\mathbb G_v(\nu_4)
\right] =
\operatorname{Tr}_{\mathrm K}
\left[
\sigma_x
\mathbb G_c(\nu_1)
\overline{\mathbb G}_h(-\nu_2)
\mathbb G_c(\nu_3)
\overline{\mathbb G}_h(-\nu_4)
\right].
\label{eqn:interband_pair_square_equivalence}
\end{equation}
The two minus signs from the valence propagators cancel, while their frequency arguments reverse under the hole transformation. The interband square is thus precisely the electron--hole pair square with the corresponding frequency routing. This equality holds before integration and spectroscopy projection, so the two representations give the same nonlinear response.

\subsection{Application to the transverse-field Ising chain}
\label{sec:tfim}

The transverse-field Ising chain provides an exactly solvable application of the bubble--square relation to a fractionalized pair continuum. We consider
\begin{equation}
 H_0=-J\sum_j\sigma_j^z\sigma_{j+1}^z
     -h\sum_j\sigma_j^x,
 \qquad
 H_{\mathrm{probe}}(t)=-b(t)M^x,
 \qquad
 M^x=\frac12\sum_j\sigma_j^x,
 \label{eqn:tfim-hamiltonian}
\end{equation}
where $\sigma_j^\alpha$ are Pauli matrices and $J,h>0$. The Jordan--Wigner transformation followed by a Bogoliubov rotation gives free fermionic quasiparticles. Following the notation and momentum convention of Wan and
Armitage~\cite{Wan2019SpinonEcho}, their dispersion and pair vertex are
\begin{equation}
 \lambda_k=2\sqrt{h^2+J^2+2hJ\cos k},
 \qquad
 f_k=\sin\theta_k=-\frac{2J\sin k}{\lambda_k}.
 \label{eqn:tfim-dispersion}
\end{equation}
Here $f_k$ is defined only for $k>0$, with each quasiparticle pair counted once. Its continuation to negative momenta is not needed when applying the bubble--square relation. Choosing the quasiparticle-pair phase so that its vertex is real, the transverse magnetization contains both scattering and pair terms,
\begin{equation}
 M^x=\sum_{k>0}\Big[
 -\cos\theta_k(n_k+n_{-k}-1)
 +\sin\theta_k
 \left(\gamma_k^\dagger\gamma_{-k}^\dagger
       +\gamma_{-k}\gamma_k\right)\Big],
 \label{eqn:tfim-magnetization}
\end{equation}
where $n_k=\gamma_k^\dagger\gamma_k$ and
$\cos\theta_k=2(h+J\cos k)/\lambda_k$.
We consider here responses per site, with $\int_{k>0}\equiv\int_0^\pi dk/(2\pi)$, so that each pair is counted once.

\subsubsection{Linear response}

The finite-frequency linear response contains two pair vertices and is therefore weighted by $\sin^2\theta_k$. Using Eq.~\eqref{eqn: bare_bubble}, with $\xi_k=\lambda_k$, its reduced pair bubble is
\begin{equation}
 \Pi_{0,\mathrm{TFIM}}^{R/A}(\omega)
 =2\int_{k>0}
 \frac{\sin^2\theta_k\,F_{-,k}}
      {\omega-2\lambda_k\pm2i\gamma},
 \qquad
 F_{-,k}=\tanh\!\left(\frac{\beta\lambda_k}{2}\right).
 \label{eqn:tfim-linear-bubble}
\end{equation}
For the real probe convention above, the physical susceptibility includes both frequency orientations,
\begin{equation}
 \chi_{xx}^{(1)}(\omega)
 =-\frac12\left[
 \Pi_{0,\mathrm{TFIM}}^R(\omega)
 +\Pi_{0,\mathrm{TFIM}}^A(-\omega)\right],
 \label{eqn:tfim-linear-frequency}
\end{equation}
where we have introduced the same notation as in Ref.~\cite{Wan2019SpinonEcho}. The scattering term in Eq.~\eqref{eqn:tfim-magnetization} is diagonal in quasiparticle occupations and does not contribute to this linear pair continuum. As above, $\gamma$ denotes a constant quasiparticle linewidth and the isolated free chain is recovered in the limit $\gamma\rightarrow0^+$.

\subsubsection{Nonlinear response}

The rephasing pathway contains four pair vertices and therefore carries the weight $\sin^4\theta_k$. Following Sec.~\ref{sec:bubble_square_relation}, we introduce
\begin{equation}
 \widetilde\Pi_{0,\mathrm{TFIM}}^{R/A}(\omega)
 =2\int_{k>0}
 \frac{\sin^4\theta_k\,F_{-,k}}
      {\omega-2\lambda_k\pm2i\gamma}.
 \label{eqn:tfim-generalized-bubble}
\end{equation}
The uniform probe conserves the momentum around the loop, so the square reduces according to Eq.~\eqref{eqn:bubble_square_relation}. Applying the ABB projection of Eq.~\eqref{eqn:ABB_echo_map} then gives
\begin{align}
 \mathcal{S}_{0,R,\mathrm{TFIM}}^R(\omega_\tau,\omega_t)
 &=i\,
 \frac{\widetilde\Pi_{0,\mathrm{TFIM}}^R(\omega_t)
       -\widetilde\Pi_{0,\mathrm{TFIM}}^A(-\omega_\tau)}
      {\omega_t+\omega_\tau+4i\gamma}=2i\int_{k>0}
 \frac{\sin^4\theta_k\,F_{-,k}}
 {(\omega_t-2\lambda_k+2i\gamma)
  (\omega_\tau+2\lambda_k+2i\gamma)}.
 \label{eqn:tfim-echo-bubble}
\end{align}
Each pair contributes at
$(\omega_\tau,\omega_t)=(-2\lambda_k,2\lambda_k)$,
forming the spinon-echo ridge. Restoring both frequency
orientations gives its real time-domain contribution,
\begin{equation}
 \chi_{xxxxx,R}^{(3)}(t,\tau)
 =-4\Theta(t)\Theta(\tau)e^{-2\gamma(t+\tau)}
 \int_{k>0}\sin^4\theta_k\,F_{-,k}
 \sin\!\left[2\lambda_k(t-\tau)\right].
 \label{eqn:tfim-echo-time-real}
\end{equation}
Here $\chi_{xxxx,R}^{(3)}$ is the ordered rephasing kernel, where the coefficient of $AB^2$ for ideal delta-like perturbations includes the additional coincident-pulse factor $1/2!$. This result agrees directly with the spinon echo obtained by Wan and Armitage~\cite{Wan2019SpinonEcho}. Following their notation, $t_1$, $t_2$, and $t_3$ denote successive coherence, waiting, and detection intervals, and their rephasing pathway is
\begin{equation}
 A_k^{(4)}
 =4\sin^4\theta_k\,
 e^{-(t_1+t_3)/T_2}e^{-t_2/T_1}
 \sin\!\left[2\lambda_k(t_3-t_1)\right].
 \label{eqn:tfim-wan-pathway}
\end{equation}
It enters the susceptibility as
$-L^{-1}\sum_{k>0}A_k^{(4)}$, with the causal step functions.
Setting $t_1=\tau$, $t_2\rightarrow0^+$, and $t_3=t$
reproduces Eq.~\eqref{eqn:tfim-echo-time-real} at zero
temperature, with $T_2^{-1}=2\gamma$. Retaining $F_{-,k}$
gives the finite-temperature weight also found in
Ref.~\cite{Spinon_Hart_2023}.

The bubble--square relation thus reproduces the known spinon echo while identifying its underlying two-point object. Its continuum envelope follows the spectral weight of $\widetilde\Pi_{0,\mathrm{TFIM}}$, weighted by $\sin^4\theta_k$, rather than the physical linear bubble, weighted by $\sin^2\theta_k$. Hence, although the nonlinear response cannot be obtained directly in terms of the linear physical susceptibility it is still given in terms of a two-point function. Particle-number-conserving vertices, proportional to $\cos\theta$, can be included through a straightforward extension of the formalism. They contribute to other response pathways but not to the rephasing echo considered here.

\subsection{Application to the twisted Kitaev chain}
\label{app:twisted_kitaev}

The twisted Kitaev chain provides a multiband extension of the previous application. We consider~\cite{Sim2023}
\begin{equation}
H_0=-J\sum_j\left[\widetilde{\sigma}_{2j-1}(\theta)\widetilde{\sigma}_{2j}(\theta)+\widetilde{\sigma}_{2j}(-\theta)\widetilde{\sigma}_{2j+1}(-\theta)\right],
\end{equation}
where $\widetilde{\sigma}_j(\theta)=\cos\theta\,\sigma_j^z+\sin\theta\,\sigma_j^y$, with the same probe $H_{\mathrm{probe}}(t)=-b(t)M^x$, $M^x=\frac12\sum_j\sigma_j^x$. The Jordan--Wigner transformation followed by a Bogoliubov rotation gives two free quasiparticle bands with positive energies $\lambda_{1,k}\geq\lambda_{2,k}$. Choosing real probe matrix elements and introducing $\Psi_k=(\gamma_{1,k},\gamma_{1,-k}^{\dagger},\gamma_{2,k},\gamma_{2,-k}^{\dagger})^T$, we have
\begin{equation}
M^x=\sum_{k>0}\Psi_k^\dagger
\begin{pmatrix}
0&c_k&s_k&0\\
c_k&0&0&s_k\\
s_k&0&0&-c_k\\
0&s_k&-c_k&0
\end{pmatrix}\Psi_k,
\qquad c_k=\cos\varphi_k,\qquad s_k=\sin\varphi_k.
\label{eq:tkm_probe}
\end{equation}
Here $\varphi_k$ is a real, momentum-dependent mixing angle parametrizing the probe matrix elements, distinct from the bond-twist angle $\theta$, so that $c_k^2+s_k^2=1$. The vertices proportional to $c_k$ create and annihilate intraband pairs, whereas those proportional to $s_k$ transfer quasiparticles between the two bands. In contrast to the diagonal scattering vertex of the TFIM, the latter changes the quasiparticle energy by $\Delta_k=\lambda_{1,k}-\lambda_{2,k}$. At zero temperature it cannot excite the quasiparticle vacuum directly, but it can act on a pair created by an earlier interaction with the field. Mixed pathways therefore survive even when the linear transfer response vanishes. We consider responses per site, with $\int_{k>0}\equiv\lim_{N\to\infty}N^{-1}\sum_{k>0}$ for a chain of $N$ sites, and count each momentum pair once. Here $k>0$ runs over the independent positive momenta in the reduced Brillouin zone of the two-site unit cell. Below we give the exact free-chain result, corresponding to the limit $\gamma\to0^+$ of the preceding subsection.

At finite temperature, define
\begin{equation}
F_{a,k}\equiv F_{-,a,k}=\tanh\left(\frac{\beta\lambda_{a,k}}{2}\right),
\qquad \Delta F_k=F_{1,k}-F_{2,k},
\qquad E_{a,k}=2\lambda_{a,k},
\qquad \Delta_k=\lambda_{1,k}-\lambda_{2,k}.
\end{equation}
With the probe vertices and common loop normalization removed, the elementary pair and transfer bubbles are
\begin{equation}
\pi_{\mathrm{p},a,k}^{R/A}(\omega)=\frac{F_{a,k}}{\omega-E_{a,k}\pm i0^+},
\qquad
\pi_{\mathrm{tr},k}^{R/A}(\omega)=\frac{\Delta F_k/2}{\omega-\Delta_k\pm i0^+}.
\label{eq:tkm_elementary_bubbles}
\end{equation}
The numerators of these bubbles have a direct occupation interpretation: $F_{a,k}=1-2n_F(\lambda_{a,k})$ is the pair-creation occupation factor, whereas $\Delta F_k/2=n_F(\lambda_{2,k})-n_F(\lambda_{1,k})$ is the occupation difference driving an interband transfer. Thus only the latter vanishes at zero temperature.

To reduce the square, we first resolve its band and particle--hole indices. In the Nambu basis above, the signed Bogoliubov energies are $(e_{1,k},e_{2,k},e_{3,k},e_{4,k})=(\lambda_{1,k},-\lambda_{1,k},\lambda_{2,k},-\lambda_{2,k})$. For $\mu,\nu\in\{1,2,3,4\}$, the scalar propagator $g_{\mu,k}(z)=(z-e_{\mu,k})^{-1}$ obeys
\begin{equation}
g_{\mu,k}(z_1)g_{\nu,k}(z_2)
=\frac{g_{\mu,k}(z_1)-g_{\nu,k}(z_2)}
{z_2-z_1+e_{\mu,k}-e_{\nu,k}}.
\label{eq:tkm_resolvent_identity}
\end{equation}
Applying this identity twice reduces each four-propagator product to two-propagator terms. The causal frequency prescriptions are retained in the complex arguments $z_j$, and coincident denominators are taken by continuity. This is the band-resolved extension of the resolvent reduction underlying Eq.~\eqref{eqn:bubble_square_relation}.

Mixed loops contain two pair vertices and two transfer vertices and therefore carry the weight $c_k^2s_k^2$. An explicit illustration is provided by the mixed two-pulse pathway $F_{2,k}\sin(E_{1,k}t+E_{2,k}\tau)$. Writing $z_t=\omega_t+i0^+$ and $z_\tau=\omega_\tau+i0^+$, its pole at positive excitation and detection frequencies is proportional to
\begin{align}
\frac{c_k^2s_k^2F_{2,k}}
{(z_t-E_{1,k})(z_\tau-E_{2,k})}
={}&\frac{c_k^2s_k^2}
{z_\tau-z_t+E_{1,k}-E_{2,k}}
\left[
\frac{F_{2,k}}{F_{1,k}}\pi_{\mathrm{p},1,k}(z_t)
-\pi_{\mathrm{p},2,k}(z_\tau)
\right],
\label{eq:tkm_mixed_bubble_example}
\end{align}
where $\pi_{\mathrm{p},a,k}(z)=F_{a,k}/(z-E_{a,k})$ denotes the analytic continuation of the corresponding pair bubble. The thermal factor follows the pair excited during $\tau$, while detection occurs at the other pair energy. Equation~\eqref{eq:tkm_mixed_bubble_example} explicitly expresses this mixed contribution in terms of two-point functions. Because the coefficient contains the momentum-dependent band splitting, it generally remains inside the momentum integral. The reduction therefore requires weighted or momentum-resolved bubbles rather than the physical linear susceptibility alone.

To obtain the full response, we evaluate the connected square with four probe matrices from Eq.~\eqref{eq:tkm_probe} and sum over the internal band and particle--hole indices. Expanding the probe vertices separates contributions containing four pair vertices, four transfer vertices, or two of each. Each band-resolved loop is reduced using Eq.~\eqref{eq:tkm_resolvent_identity}, retaining its thermal occupation factors. Performing the ABB projection, with successive coherence, waiting, and detection intervals $t_1=\tau$, $t_2\to0^+$, and $t_3=t$, and collecting terms gives
\begin{align}
\chi_{xxxx,\mathrm{TKM}}^{(3)}(t,\tau)
={}&-4\Theta(t)\Theta(\tau)\int_{k>0}\Bigg\{
\sum_{a=1,2}F_{a,k}\left[c_k^2\sin\bigl(E_{a,k}(t+\tau)\bigr)+c_k^4\sin\bigl(E_{a,k}(t-\tau)\bigr)\right]\nonumber\\
&\qquad+\Delta F_k\left[s_k^2\sin\bigl(\Delta_k(t+\tau)\bigr)+s_k^2(s_k^2-c_k^2)\sin\bigl(\Delta_k(t-\tau)\bigr)\right]
+c_k^2s_k^2\mathcal{M}_k(t,\tau)\Bigg\},
\label{eq:tkm_abb}
\end{align}
where the off-diagonal mixed pathways are
\begin{multline}
\mathcal{M}_k(t,\tau)
=F_{2,k}\sin(E_{1,k}t+E_{2,k}\tau)+F_{1,k}\sin(E_{2,k}t+E_{1,k}\tau)
-2F_{2,k}\sin(\Delta_k t+E_{2,k}\tau)-2F_{1,k}\sin(-\Delta_k t+E_{1,k}\tau)\\
+\Delta F_k\left[\sin(E_{1,k}t-\Delta_k\tau)-\sin(E_{2,k}t+\Delta_k\tau)\right].
\label{eq:tkm_mixed}
\end{multline}
The mixed pathways proportional to $c_k^2s_k^2$ account for the off-diagonal features identified by Sim et al.~\cite{Sim2023}, which distinguish the twisted Kitaev chain from the TFIM despite their similar linear spectra. The present construction expresses these pathways through weighted, band-resolved two-point functions, illustrating how nonlinear spectroscopy can reveal information beyond the physical linear susceptibility without requiring independent higher-order correlations of the constituent quasiparticles.

The first four sine terms in $\mathcal{M}_k$ describe a pair excitation during $\tau$, followed by detection at the other pair energy or at the interband difference. The final two terms, proportional to $\Delta F_k$, describe an interband excitation during $\tau$ followed by pair detection; these require thermal quasiparticles. For example, $\sin(E_{1,k}t+E_{2,k}\tau)$ contributes at $(\omega_\tau,\omega_t)=(E_{2,k},E_{1,k})$, together with its frequency-inverted partner. In particular, the terms independent of $\Delta F_k$ retain off-diagonal peaks at zero temperature.

The coefficient of $AB^2$ for ideal delta-like perturbations includes the additional coincident-pulse factor $1/2!$, as in the preceding application. The pair echo follows from the generalized bubble
\begin{equation}
\widetilde{\Pi}_{0,\mathrm{TKM}}^{R/A}(\omega)=2\int_{k>0}c_k^4\sum_{a=1,2}\pi_{\mathrm{p},a,k}^{R/A}(\omega).
\end{equation}
Each branch contributes along $\omega_t=-\omega_\tau$ at the pair energy $E_{a,k}$. At finite temperature, transfer processes also support rephasing at $\Delta_k$, with a total weight $\Delta F_k s_k^2(s_k^2-c_k^2)$ arising from interference between pure-transfer and mixed pathways. The remaining terms correlate different excitation and detection frequencies and generate off-diagonal features. In the zero-temperature limit, $F_{a,k}\to1$ and $\Delta F_k\to0$, Eq.~\eqref{eq:tkm_abb} reproduces the complete ordered ABB susceptibility $\chi_{xxxx}^{(3,2)}(t,t,\tau+t)$ of Sim et al.~\cite{Sim2023}. Identifying $\lambda_{1,k}=\lambda_k$ and $\lambda_{2,k}=l_k$, the nonrephasing and rephasing terms in the band sum reproduce their pathways $Q_k^{(1)}$ and $Q_k^{(2)}$, while the first two and following two sine terms of $\mathcal{M}_k$ reproduce $Q_k^{(3)}$ and $Q_k^{(4)}$, respectively. The thermal factors $F_{a,k}$ and the terms proportional to $\Delta F_k$ provide the corresponding finite-temperature generalization.
% The construction therefore extends the bubble--square reduction to thermally occupied multiband continua, while retaining the distinction between generalized two-point functions and the measured linear susceptibility.

\section{Semicircular pair continuum}
\label{app:semicircular_model}

In this Appendix we collect the analytical results for the semicircular pair continuum used in the figures throughout the main text. The purpose of this model is to provide a simple continuum for which both the bare pair bubble and its ladder resummation can be evaluated analytically. In particular, the model allows us to follow explicitly the emergence of a bound state below (above) the continuum as the attractive (repulsive) interaction is increased.

\subsection{Bare pair bubble}
\label{app:semicircular_bare}

We begin from the noninteracting pair bubble introduced in Eq.~\eqref{eqn: bare_bubble},
\begin{equation}
\Pi_0^{R/A}(\omega)=2\int_{\mathbf{k}}\frac{F_{\eta,\mathbf{k}}}{\omega-2\xi_{\mathbf{k}}\pm 2i\gamma}.
\end{equation}
For the purposes of the toy model, we absorb the smooth equilibrium
factor $F_{\eta,\mathbf{k}}$, together with the overall normalization
of the reduced loop, into an effective pair density of states. We take
this density of states to be semicircular,
\begin{equation}
\rho_{\rm sc}(\epsilon)
=
\frac{2}{\pi \delta\omega_0^2}
\sqrt{
\delta\omega_0^2-(\epsilon-\omega_0)^2
}
\,
\Theta
\left(
\delta\omega_0-|\epsilon-\omega_0|
\right), \qquad \text{with} \qquad \int_{-\infty}^{\infty}
d\epsilon\,
\rho_{\rm sc}(\epsilon)
=
1.
\label{eqn:semicircular_dos}
\end{equation}
where $\omega_0$ denotes the center of the pair continuum and $\delta\omega_0$ its
half bandwidth. We have chosen the normalization to be unity, but an arbitrary overall spectral weight can be restored by multiplying $\rho_{\rm sc}$ and $\Pi$ by the same constant. The corresponding retarded and advanced pair bubbles are
\begin{equation}
\Pi_0^{R/A}(\omega)
=
\int_{\omega_0-\delta\omega_0}^{\omega_0+\delta\omega_0}
d\epsilon\,
\frac{
\rho_{\rm sc}(\epsilon)
}{
\omega-\epsilon\pm 2i\gamma
}.
\label{eqn:semicircular_bubble_integral}
\end{equation}
Introducing $z_{R/A}=\omega-\omega_0\pm 2i\gamma$, and shifting the integration variable according to
$x=\epsilon-\omega_0$, Eq.~\eqref{eqn:semicircular_bubble_integral}
becomes
\begin{equation}
\Pi_0^{R/A}(\omega)
=
\frac{2}{\pi \delta\omega_0^2}
\int_{-\delta\omega_0}^{\delta\omega_0}
dx\,
\frac{
\sqrt{\delta\omega_0^2-x^2}
}{
z_{R/A}-x
}=
\frac{2}{\delta\omega_0^2}
\left[
z_{R/A}
-
\sqrt{z_{R/A}-\delta\omega_0}
\sqrt{z_{R/A}+\delta\omega_0}
\right].
\label{eqn:semicircular_pi}
\end{equation}
The square roots in Eq.~\eqref{eqn:semicircular_pi} are taken on
their principal branches. Written as the product of the two square roots,
this prescription automatically gives the retarded branch for
$\operatorname{Im}z>0$ and the advanced branch for
$\operatorname{Im}z<0$. 

\subsection{RPA-dressed pair bubble}
\label{app:semicircular_rpa}

We now include the interaction $H_{\rm int}=gP^\dagger P$, using the convention of the main text, such that $g<0$ corresponds to
attraction and $g>0$ to repulsion. Within the particle--particle ladder
approximation, the dressed pair bubble is
\begin{equation}
\Pi^{R/A}(\omega)
=
\frac{
\Pi_0^{R/A}(\omega)
}{
1-g\Pi_0^{R/A}(\omega)
},
\label{eqn:semicircular_dressed_bubble}
\end{equation}
cf. Eq.~\eqref{eqn:resummed_pair_bubble}, and the corresponding ladder vertex is
\begin{equation}
\mathcal{L}^{R/A}(\omega)
=
\frac{
1
}{
1-g\Pi_0^{R/A}(\omega)
}.
\label{eqn:semicircular_ladder}
\end{equation}

The collective pole of the dressed pair propagator is determined by
\begin{equation}
1-g\Pi_0^R(\omega_b)=0.
\label{eqn:semicircular_pole_condition}
\end{equation}
We first consider an attractive interaction and write $g=-\abs{g}$. A bound state must lie below the lower continuum edge, $\omega_b<\omega_0-\delta\omega_0$, where $\Pi_0^R(\omega)$ is real and negative. The pole condition becomes $1+\abs{g}\Pi_0^R(\omega_b)=0$. The pole position can be obtained analytically. Defining $x_b=\omega_b-\omega_0$, the pole condition reads
\begin{equation}
1-\frac{2g}{\delta\omega_0^2}\left[x_b+\sqrt{x_b^2-\delta\omega_0^2}\right]=0
\end{equation}
for the attractive solution below the band, and has a solution only if $\abs{g}\geq \delta\omega_0/2$. Solving this equation gives the binding energy equation
\begin{equation}
\omega_b=\omega_0-\abs{g}-\frac{\delta\omega_0^2}{4\abs{g}}.
\label{eqn:semicircular_bound_state_energy}
\end{equation}
The semicircular model treats attractive and repulsive bound states
symmetrically. Attraction generates a pole below the continuum, whereas
sufficiently strong repulsion generates, analogously, a pole above it. We focus here on the attractive case only.

The spectral weight transferred into the collective mode can also be
obtained analytically. Expanding Eq.~\eqref{eqn:semicircular_dressed_bubble}
around the pole,
\begin{equation}
\Pi^R(\omega)\simeq\frac{Z_b}{\omega-\omega_b+i0^+}+\Pi_{\rm reg}^R(\omega),
\qquad
\text{with}
\qquad 
Z_b=\frac{Z_{\mathcal{D}}}{g^2}=-\frac{1}{g^2\partial_\omega\Pi_0^R(\omega_b)}.
\label{eqn:semicircular_pole_expansion}
\end{equation}
For the semicircular bubble,
\begin{equation}
\partial_\omega\Pi_0^R(\omega)
=
\frac{2}{\delta\omega_0^2}
\left[
1
+
\frac{
\omega-\omega_0
}{
\sqrt{(\omega-\omega_0)^2-\delta\omega_0^2}
}
\right],\qquad Z_b=1
-
\frac{
\delta\omega_0^2
}{
4g^2
},
\end{equation}
where in the second equality we have used Eq.~\eqref{eqn:semicircular_bound_state_energy}. The residue vanishes continuously at the threshold,
\begin{equation}
Z_b
\rightarrow 0,
\qquad
|g|\rightarrow\frac{\delta\omega_0}{2},
\end{equation}
and approaches unity at strong coupling,
\begin{equation}
Z_b
\rightarrow 1,
\qquad
|g|\gg \delta\omega_0.
\end{equation}

Since the normalized bare bubble behaves as $\Pi_0(z)\simeq1/z$ at large frequency, the dressed bubble has the same leading asymptotic behavior, $\Pi(z)
\simeq 1/z$. The total spectral weight is therefore conserved. Once the isolated pole
has detached from the continuum, the remaining continuum weight is
\begin{equation}
Z_{\rm cont}
=
1-Z_b
=
\frac{
\delta\omega_0^2
}{
4g^2
}.
\label{eqn:semicircular_continuum_weight}
\end{equation}
The strong-coupling limit is therefore particularly transparent: the
spectral weight of the continuum vanishes as $\delta\omega_0^2/g^2$, while the bound state asymptotically carries the full pair spectral weight.

\newpage

%%%%%%%%%%%%%%%%%%%%%%%%%%%%%%%%%%%%%%%%%%%%%%%

\bibliographystyle{elsarticle-num}
\bibliography{Bib}

@misc{salvador2024principles,
      title={Principles of 2D terahertz spectroscopy of collective excitations: the case of Josephson plasmons in layered superconductors}, 
      author={Alex Gómez Salvador and Pavel E. Dolgirev and Marios H. Michael and Albert Liu and Danica Pavicevic and Michael Fechner and Andrea Cavalleri and Eugene Demler},
      year={2024},
      eprint={2401.05503},
      archivePrefix={arXiv},
      primaryClass={cond-mat.supr-con}
}

@Article{Cundiff2013,
author={Cundiff, Steven T.
and Mukamel, Shaul},
title={Optical multidimensional coherent spectroscopy},
journal={Physics Today},
year={2013},
month={Jul},
day={01},
volume={66},
number={7},
pages={44-49},
issn={0031-9228},
doi={10.1063/PT.3.2047},
url={https://doi.org/10.1063/PT.3.2047}
}

@book{hamm_zanni_2011, place={Cambridge}, title={Concepts and Methods of 2D Infrared Spectroscopy}, DOI={10.1017/CBO9780511675935}, publisher={Cambridge University Press}, author={Hamm, Peter and Zanni, Martin}, year={2011}}

@book{Mukamel,
  title={Principles of Nonlinear Optical Spectroscopy},
  author={Mukamel, S.},
  isbn={9780195092783},
  lccn={gb95087868},
  series={Oxford series in optical and imaging sciences},
  url={https://books.google.ch/books?id=k\_7uAAAAMAAJ},
  year={1995},
  publisher={Oxford University Press}
}

@book{kamenev2023field,
  title={Field theory of non-equilibrium systems},
  author={Kamenev, Alex},
  year={2023},
  publisher={Cambridge University Press}
}

@article{Liu_2023_echo,
author={Liu, A.
and Pavi{\'{c}}evi{\'{c}}, D.
and Michael, M. H.
and Salvador, A. G.
and Dolgirev, P. E.
and Fechner, M.
and Disa, A. S.
and M. Lozano, P.
and Li, Q.
and Gu, G. D.
and Demler, E.
and Cavalleri, A.},
title={Probing inhomogeneous cuprate superconductivity by terahertz Josephson echo spectroscopy},
journal={Nature Physics},
year={2024},
month={Nov},
day={01},
volume={20},
number={11},
pages={1751-1756},
issn={1745-2481},
doi={10.1038/s41567-024-02643-5},
url={https://doi.org/10.1038/s41567-024-02643-5}
}

@article{Hahn_1950,
  title = {Spin Echoes},
  author = {Hahn, E. L.},
  journal = {Phys. Rev.},
  volume = {80},
  issue = {4},
  pages = {580--594},
  numpages = {0},
  year = {1950},
  month = {Nov},
  publisher = {American Physical Society},
  doi = {10.1103/PhysRev.80.580},
  url = {https://link.aps.org/doi/10.1103/PhysRev.80.580}
}

@article{Spinon_Hart_2023,
  title = {Extracting spinon self-energies from two-dimensional coherent spectroscopy},
  author = {Hart, Oliver and Nandkishore, Rahul},
  journal = {Phys. Rev. B},
  volume = {107},
  issue = {20},
  pages = {205143},
  numpages = {20},
  year = {2023},
  month = {May},
  publisher = {American Physical Society},
  doi = {10.1103/PhysRevB.107.205143},
  url = {https://link.aps.org/doi/10.1103/PhysRevB.107.205143}
}

@article{fava2023divergent,
  title = {Divergent Nonlinear Response from Quasiparticle Interactions},
  author = {Fava, Michele and Gopalakrishnan, Sarang and Vasseur, Romain and Essler, Fabian and Parameswaran, S. A.},
  journal = {Phys. Rev. Lett.},
  volume = {131},
  issue = {25},
  pages = {256505},
  numpages = {7},
  year = {2023},
  month = {Dec},
  publisher = {American Physical Society},
  doi = {10.1103/PhysRevLett.131.256505},
  url = {https://link.aps.org/doi/10.1103/PhysRevLett.131.256505}
}

@article{mcginley2022signatures,
  title = {Signatures of Fractional Statistics in Nonlinear Pump-Probe Spectroscopy},
  author = {McGinley, Max and Fava, Michele and Parameswaran, S. A.},
  journal = {Phys. Rev. Lett.},
  volume = {132},
  issue = {6},
  pages = {066702},
  numpages = {7},
  year = {2024},
  month = {Feb},
  publisher = {American Physical Society},
  doi = {10.1103/PhysRevLett.132.066702},
  url = {https://link.aps.org/doi/10.1103/PhysRevLett.132.066702}
}

@article{parameswaran2020asymptotically1,
  title = {Asymptotically Exact Theory for Nonlinear Spectroscopy of Random Quantum Magnets},
  author = {Parameswaran, S. A. and Gopalakrishnan, S.},
  journal = {Phys. Rev. Lett.},
  volume = {125},
  issue = {23},
  pages = {237601},
  numpages = {6},
  year = {2020},
  month = {Dec},
  publisher = {American Physical Society},
  doi = {10.1103/PhysRevLett.125.237601},
  url = {https://link.aps.org/doi/10.1103/PhysRevLett.125.237601}
}

@Article{Liu2025Qmat,
author={Liu, Albert},
title={Multidimensional terahertz probes of quantum materials},
journal={npj Quantum Materials},
year={2025},
month={Feb},
day={10},
volume={10},
number={1},
pages={18},
issn={2397-4648},
doi={10.1038/s41535-025-00741-y},
url={https://doi.org/10.1038/s41535-025-00741-y}
}

@article{fava2024longtimedivergencesnonlinearresponse,
title = {Long-time divergences in the nonlinear response of gapped one-dimensional many-particle systems},
pages = {086},
author = {Fava, Michele and Gopalakrishnan, Sarang and Vasseur, Romain and Parameswaran, Siddharth and Essler, Fabian H. L.},
journal = {SciPost Phys.},
volume = {19},
year = {2025},
publisher = {SciPost},
doi = {10.21468/SciPostPhys.19.4.086},
url = {https://scipost.org/10.21468/SciPostPhys.19.4.086}
}

@article{salvador2025Echo_theory,
  title = {Two-dimensional spectroscopy of bosonic collective excitations in disordered many-body systems},
  author = {G{\'o}mez-Salvador, Alex and Morera, Ivan and Michael, Marios H. and Dolgirev, Pavel E. and Pavicevic, Danica and Liu, Albert and Cavalleri, Andrea and Demler, Eugene},
  journal = {Phys. Rev. Res.},
  volume = {8},
  issue = {4},
  pages = {043031},
  numpages = {22},
  year = {2026},
  month = {Oct},
  publisher = {American Physical Society},
  doi = {10.1103/zsbw-2yzb},
  url = {https://link.aps.org/doi/10.1103/zsbw-2yzb}
}

@misc{huang2025review,
      title={Unlocking Quantum Control and Multi-Order Correlations via Terahertz Two-Dimensional Coherent Spectroscopy}, 
      author={Chuankun Huang and Martin Mootz and Liang Luo and Ilias E. Perakis and Jigang Wang},
      year={2025},
      eprint={2507.02116},
      archivePrefix={arXiv},
      primaryClass={cond-mat.supr-con},
      url={https://arxiv.org/abs/2507.02116}, 
}

@misc{salvador2025gapinhomogeneitycupratesview,
      title={Gap Inhomogeneity in Cuprates: a view from Two-Dimensional Josephson Echo Spectroscopy}, 
      author={Alex Gómez Salvador and Ivan Morera and Marios H. Michael and Pavel E. Dolgirev and Danica Pavicevic and Albert Liu and Andrea Cavalleri and Eugene Demler},
      year={2025},
      eprint={2509.23856},
      archivePrefix={arXiv},
      primaryClass={cond-mat.supr-con},
      url={https://arxiv.org/abs/2509.23856}, 
}

@article{fiore_2026_2D_diagrams,
  title = {Two-dimensional terahertz spectroscopy in electronic systems: A many-body diagrammatic approach},
  author = {Fiore, Jacopo and Sellati, Niccol\`o and Udina, Mattia and Benfatto, Lara},
  journal = {Phys. Rev. B},
  volume = {113},
  issue = {17},
  pages = {174524},
  numpages = {26},
  year = {2026},
  month = {May},
  publisher = {American Physical Society},
  doi = {10.1103/8ffz-rtyf},
  url = {https://link.aps.org/doi/10.1103/8ffz-rtyf}
}

@Article{Huang2026,
author={Huang, Chuankun
and Mootz, Martin
and Luo, Liang
and Perakis, Ilias E.
and Wang, Jigang},
title={Terahertz 2D coherent spectroscopy for probing and controlling multicorrelations in quantum matter},
journal={Nature Reviews Physics},
year={2026},
month={Mar},
day={01},
volume={8},
number={3},
pages={171-185},
issn={2522-5820},
doi={10.1038/s42254-025-00917-2},
url={https://doi.org/10.1038/s42254-025-00917-2}
}

@article{Wan2019SpinonEcho,
  title = {Resolving Continua of Fractional Excitations by Spinon Echo in {THz} 2D Coherent Spectroscopy},
  author = {Wan, Yuan and Armitage, N. P.},
  journal = {Physical Review Letters},
  volume = {122},
  number = {25},
  pages = {257401},
  year = {2019},
  doi = {10.1103/PhysRevLett.122.257401}
}

@article{SmallwoodCundiff2018,
author = {Smallwood, Christopher L. and Cundiff, Steven T.},
title = {Multidimensional Coherent Spectroscopy of Semiconductors},
journal = {Laser \& Photonics Reviews},
volume = {12},
number = {12},
pages = {1800171},
doi = {https://doi.org/10.1002/lpor.201800171},
url = {https://onlinelibrary.wiley.com/doi/abs/10.1002/lpor.201800171},
eprint = {https://onlinelibrary.wiley.com/doi/pdf/10.1002/lpor.201800171},
year = {2018}
}

@article{Choi_2020,
  title = {Theory of Two-Dimensional Nonlinear Spectroscopy for the Kitaev Spin Liquid},
  author = {Choi, Wonjune and Lee, Ki Hoon and Kim, Yong Baek},
  journal = {Phys. Rev. Lett.},
  volume = {124},
  issue = {11},
  pages = {117205},
  numpages = {5},
  year = {2020},
  month = {Mar},
  publisher = {American Physical Society},
  doi = {10.1103/PhysRevLett.124.117205},
  url = {https://link.aps.org/doi/10.1103/PhysRevLett.124.117205}
}

@article{Gao_2023,
  title = {Two-dimensional coherent spectrum of interacting spinons from matrix product states},
  author = {Gao, Qi and Liu, Yang and Liao, Haijun and Wan, Yuan},
  journal = {Phys. Rev. B},
  volume = {107},
  issue = {16},
  pages = {165121},
  numpages = {8},
  year = {2023},
  month = {Apr},
  publisher = {American Physical Society},
  doi = {10.1103/PhysRevB.107.165121},
  url = {https://link.aps.org/doi/10.1103/PhysRevB.107.165121}
}

@article{Li_2021,
  author  = {Li, Zi-Long and Oshikawa, Masaki and Wan, Yuan},
  title   = {Photon Echo from Lensing of Fractional Excitations in Tomonaga--Luttinger Spin Liquid},
  journal = {Phys. Rev. X},
  volume  = {11},
  pages   = {031035},
  year    = {2021},
  doi     = {10.1103/PhysRevX.11.031035}
}

@article{Watanabe_2024,
  title = {Exploring two-dimensional coherent spectroscopy with exact diagonalization: Spinons and confinement in one-dimensional quantum magnets},
  author = {Watanabe, Yoshito and Trebst, Simon and Hickey, Ciar\'an},
  journal = {Phys. Rev. B},
  volume = {110},
  issue = {13},
  pages = {134443},
  numpages = {23},
  year = {2024},
  month = {Oct},
  publisher = {American Physical Society},
  doi = {10.1103/PhysRevB.110.134443},
  url = {https://link.aps.org/doi/10.1103/PhysRevB.110.134443}
}

@article{Potts_2024,
  title = {Signatures of Spinon Dynamics and Phase Structure of Dipolar-Octupolar Quantum Spin Ices in Two-Dimensional Coherent Spectroscopy},
  author = {Potts, Mark and Moessner, Roderich and Benton, Owen},
  journal = {Phys. Rev. Lett.},
  volume = {133},
  issue = {22},
  pages = {226701},
  numpages = {6},
  year = {2024},
  month = {Nov},
  publisher = {American Physical Society},
  doi = {10.1103/PhysRevLett.133.226701},
  url = {https://link.aps.org/doi/10.1103/PhysRevLett.133.226701}
}

@article{Qiang_2024,
  title = {Probing Majorana Wave Functions in Kitaev Honeycomb Spin Liquids with Second-Order Two-Dimensional Spectroscopy},
  author = {Qiang, Yihua and Quito, Victor L. and Trevisan, Tha\'{\i}s V. and Orth, Peter P.},
  journal = {Phys. Rev. Lett.},
  volume = {133},
  issue = {12},
  pages = {126505},
  numpages = {6},
  year = {2024},
  month = {Sep},
  publisher = {American Physical Society},
  doi = {10.1103/PhysRevLett.133.126505},
  url = {https://link.aps.org/doi/10.1103/PhysRevLett.133.126505}
}

@misc{kirchner2025measuringanyonicexchangephases,
      title={Measuring Anyonic Exchange Phases Using Two-Dimensional Coherent Spectroscopy}, 
      author={Nico Kirchner and Wonjune Choi and Frank Pollmann},
      year={2025},
      eprint={2511.17420},
      archivePrefix={arXiv},
      primaryClass={cond-mat.str-el},
      url={https://arxiv.org/abs/2511.17420}, 
}

@article{FleuryLoudon1968,
  author  = {Fleury, P. A. and Loudon, R.},
  title   = {Scattering of Light by One- and Two-Magnon Excitations},
  journal = {Phys. Rev.},
  volume  = {166},
  pages   = {514},
  year    = {1968},
  doi     = {10.1103/PhysRev.166.514}
}

@article{Perreault2015,
  author  = {Perreault, Brent and Knolle, Johannes
             and Perkins, Natalia B. and Burnell, F. J.},
  title   = {Theory of {Raman} response in three-dimensional
             {Kitaev} spin liquids: Application to
             $\beta$- and $\gamma$-$\mathrm{Li}_{2}\mathrm{IrO}_{3}$
             compounds},
  journal = {Phys. Rev. B},
  volume  = {92},
  pages   = {094439},
  year    = {2015},
  doi     = {10.1103/PhysRevB.92.094439}
}

@article{Potts2024,
  author  = {Potts, Mark and Moessner, Roderich and Benton, Owen},
  title   = {Exploiting polarization dependence in two-dimensional
             coherent spectroscopy: Examples of
             $\mathrm{Ce}_{2}\mathrm{Zr}_{2}\mathrm{O}_{7}$ and
             $\mathrm{Nd}_{2}\mathrm{Zr}_{2}\mathrm{O}_{7}$},
  journal = {Phys. Rev. B},
  volume  = {109},
  pages   = {104435},
  year    = {2024},
  doi     = {10.1103/PhysRevB.109.104435}
}

@article{Sim2023,
   author = {Sim, GiBaik and Pollmann, Frank and Knolle, Johannes},
   title = {Microscopic details of two-dimensional spectroscopy of one-dimensional quantum Ising magnets},
   journal = {Physical Review B},
   volume = {108},
   pages = {134423},
   year = {2023},
   doi = {10.1103/PhysRevB.108.134423}
 }

@article{Morampudi_2017,
  title = {Statistics of Fractionalized Excitations through Threshold Spectroscopy},
  author = {Morampudi, Siddhardh C. and Turner, Ari M. and Pollmann, Frank and Wilczek, Frank},
  journal = {Phys. Rev. Lett.},
  volume = {118},
  issue = {22},
  pages = {227201},
  numpages = {5},
  year = {2017},
  month = {May},
  publisher = {American Physical Society},
  doi = {10.1103/PhysRevLett.118.227201},
  url = {https://link.aps.org/doi/10.1103/PhysRevLett.118.227201}
}

@book{Mahan2000,
  author    = {Mahan, Gerald D.},
  title     = {Many-Particle Physics},
  edition   = {3},
  publisher = {Kluwer Academic/Plenum Publishers},
  year      = {2000},
  doi       = {10.1007/978-1-4757-5714-9}
}

@book{Mukamel1995,
  author    = {Mukamel, Shaul},
  title     = {Principles of Nonlinear Optical Spectroscopy},
  publisher = {Oxford University Press},
  year      = {1995}
}

@article{KamenevLevchenko2009,
  author  = {Kamenev, Alex and Levchenko, Alex},
  title   = {Keldysh technique and non-linear sigma-model:
             basic principles and applications},
  journal = {Advances in Physics},
  volume  = {58},
  number  = {3},
  pages   = {197--319},
  year    = {2009},
  doi     = {10.1080/00018730902850504}
}

@article{KotliarLiu1988,
  author  = {Kotliar, Gabriel and Liu, Jialin},
  title   = {Superexchange mechanism and d-wave superconductivity},
  journal = {Phys. Rev. B},
  volume  = {38},
  pages   = {5142--5145},
  year    = {1988},
  doi     = {10.1103/PhysRevB.38.5142}
}

@article{LeeNagaosaWen2006,
  author  = {Lee, Patrick A. and Nagaosa, Naoto and Wen, Xiao-Gang},
  title   = {Doping a Mott insulator: Physics of high-temperature superconductivity},
  journal = {Rev. Mod. Phys.},
  volume  = {78},
  pages   = {17--85},
  year    = {2006},
  doi     = {10.1103/RevModPhys.78.17}
}

@article{Scalapino2012,
  author  = {Scalapino, D. J.},
  title   = {A common thread: The pairing interaction for unconventional superconductors},
  journal = {Rev. Mod. Phys.},
  volume  = {84},
  pages   = {1383--1417},
  year    = {2012},
  doi     = {10.1103/RevModPhys.84.1383}
}

@misc{andrei2026universalmagneticenergyscale,
      title={Universal magnetic energy scale in the doped Fermi-Hubbard model}, 
      author={Radu Andrei and Ivan Morera and Jonathan B. Curtis and Immanuel Bloch and Eugene Demler},
      year={2026},
      eprint={2604.15234},
      archivePrefix={arXiv},
      primaryClass={cond-mat.str-el},
      url={https://arxiv.org/abs/2604.15234}, 
}

@misc{liu2026probingphasefluctuationscuprates,
      title={Probing Phase Fluctuations in Cuprates by Multidimensional Josephson Spectroscopy}, 
      author={Albert Liu and Danica Pavicevic and Alex Gómez Salvador and Marios H. Michael and Michael Fechner and Pedro M. Lozano and Qiang Li and Genda D. Gu and Eugene Demler and Andrea Cavalleri},
      year={2026},
      eprint={2611.21708},
      archivePrefix={arXiv},
      primaryClass={cond-mat.supr-con},
      url={https://arxiv.org/abs/2611.21708}, 
}

%%%%%%%%%%%%%%%%%%%%%%%%%%%%%%%%%%%%%%%%%%%%%%%

\end{document}